\documentclass[pra,superscriptaddress,twocolumn,amsmath,amssymb,aps]{revtex4-1}

\usepackage{graphicx} 
\usepackage{dcolumn} 
\usepackage{bm} 
\usepackage{braket}
\usepackage{color}
\usepackage{amsthm}
\usepackage{hyperref} 
\usepackage[normalem]{ulem} 
\usepackage{qcircuit}       

\definecolor{lightblue}{RGB}{73,151,208}
\definecolor{crimson}{RGB}{140,41,53}

\hypersetup{
    colorlinks,
    linkcolor={crimson},
    citecolor={lightblue},
    urlcolor={lightblue}
}

\newtheorem{definition}{Definition}

\newtheorem{theorem}{Theorem}

\def\>{\rangle}
\def\<{\langle}
\newcommand{\state}[1]{\left \vert \left. #1 \right\rangle   \right.}
\newcommand{\bstate}[1]{\left\langle  \left.  #1  \right \vert \right. }

\newcommand{\sho}[2]{\textcolor{red}{#1 {\sout {#2}}}}

\usepackage{mathtools}

\begin{document}

\preprint{APS/123-QED}

\title{Power of sequential protocols in hidden quantum channel discrimination} 

\author{Sho Sugiura}\affiliation{
Physics and Informatics Laboratory, NTT Research, Inc.,940 Stewart Dr., Sunnyvale, California, 94085, USA } \affiliation{Laboratory for Nuclear Science, Massachusetts Institute of Technology, Cambridge, Massachusetts 02139, USA}
\author{Arkopal Dutt}\affiliation{Department of Physics, Co-Design Center for Quantum Advantage, Massachusetts Institute of Technology, Cambridge, Massachusetts 02139, USA}
\author{William J. Munro}
\affiliation{NTT Basic Research Laboratories and Research Center for Theoretical Quantum Physics,3-1 Morinosato-Wakamiya, Atsugi, Kanagawa 243-0198, Japan}
\author{Sina Zeytino\u{g}lu}\affiliation{
Physics and Informatics Laboratory, NTT Research, Inc.,940 Stewart Dr., Sunnyvale, California, 94085, USA } \affiliation{Department of Physics, Harvard University, Cambridge, Massachusetts 02138, USA}
\author{Isaac L. Chuang}\affiliation{Department of Physics, Co-Design Center for Quantum Advantage, Massachusetts Institute of Technology, Cambridge, Massachusetts 02139, USA} \affiliation{Department of Electrical Engineering and Computer Science, Massachusetts Institute of Technology, Cambridge, Massachusetts 02139, USA}
\date{\today}%

\begin{abstract}
In many natural and engineered systems, unknown quantum channels act on a subsystem that cannot be directly controlled and measured, but is instead learned through a controllable subsystem that weakly interacts with it. We study quantum channel discrimination (QCD) under these restrictions, which we call hidden system QCD (HQCD). We find that sequential protocols achieve perfect discrimination and saturate the Heisenberg limit. In contrast, depth-1 parallel and multi-shot protocols cannot solve HQCD. This suggests that sequential protocols are superior in experimentally realistic situations.
\end{abstract}

\maketitle

\begin{figure}[b]
    \centering
    \includegraphics[width=0.95\linewidth]{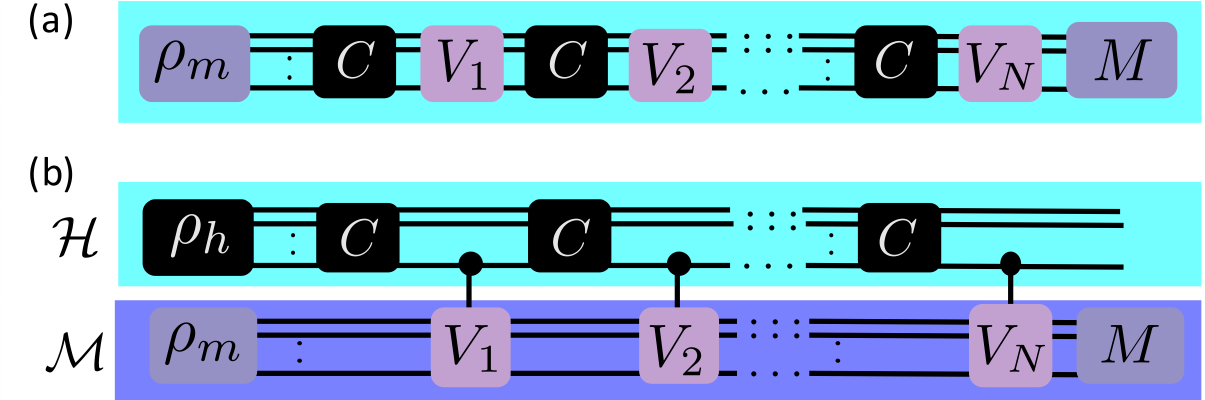}
    \caption{
    Comparison between conventional quantum channel discrimination (QCD) and hidden quantum channel discrimination (HQCD). Here the black boxes indicate the unknown channels and state. In both cases, the action of unknown channel $C$ is inferred by selecting an input state $\rho_m$, applying controlled $V_n$ operations ($n = 1,\ldots,N$), and measuring with $M$ to minimize the error probability. (a) Conventional QCD, involving the direct manipulation and measurement of the system. (b) HQCD, where the physical system $\mathcal{H}$ and measurement system $\mathcal{M}$ are explicitly distinguished.}
    \label{fig:channel estimation}
\end{figure}

{\it Introduction: }
Discriminating between physical operations, often called quantum channel discrimination (QCD) in quantum information science, is a fundamental task in experiments \cite{acin,Duan98,Calsamiglia08.77.032311,zhuang_pirandola,pirandola,Wang19}. In QCD, an unknown physical operation is modeled as a quantum channel $C$ through a completely-positive trace-preserving map acting on the system of interest \cite{nielsen_chuang}. The goal is to identify $C$ from known alternatives using a discrimination protocol. Discrimination protocols are considered efficient (1) when a desired error probability is achieved with fewer queries compared to classical methods \cite{Vittorio04,Braun18} or particularly successful (2) when the error probability is zero 
\cite{acin,Duan98, Duan}. For example, sequential protocols \cite{Duan98} involve an initial state $\rho_m$, and a positive operator-valued measurement (POVM) $M$, and $N$ queries, each consisting of the unknown channel $C$ and tunable unitary operations $V_n$ ($n=1,\ldots,N$), as shown in Fig.~\ref{fig:channel estimation}(a). Protocols including sequential and parallel protocols are able to achieve (1) and (2) when arbitrary operations of $V_n$ and measurements $M$ are allowed on the system \cite{Harrow10,Pezze18, Braun18}. 

While conventional QCD considers a fully controllable system, experimental systems often consist of a fully-controllable subsystem, which we call the measurement system $\mathcal{M}$, and an uncontrollable subsystem, which we call the channel system $\mathcal{H}$ \cite{Imoto85,Philippe98,Schmidt15, Or22,Pechal21,xiang2013hybrid}. Here, $\mathcal{M}$ interacts with $\mathcal{H}$ to detect the action of $C$ on $\mathcal{H}$. Such composite systems are used in quantum non-demolition measurements \cite{Imoto85, Philippe98}, quantum logic detection \cite{Schmidt15, Or22}, and occur in designs of superconducting quantum devices \cite{Pechal21}. 

These experiments motivate us to consider the following restrictions on system $\mathcal{H}$ in QCD: arbitrary control of $\mathcal{H}$ is not possible, measurement on $\mathcal{H}$ is not allowed, and the initialization of $\mathcal{H}$ is unreliable. 
The state on $\mathcal{H}$ thus evolves only under the dynamics $C$ on $\mathcal{H}$. The separation between $\mathcal{H}$ and $\mathcal{M}$ motivates the third restriction as one no longer has control over the state preparation on $\mathcal{H}$ and the initial state cannot be purified. We call $\mathcal{H}$ probed under these three restrictions hidden, and the associated channel discrimination problem Hidden system Quantum Channel Discrimination (HQCD). The effect of these restrictions on a conventional sequential QCD protocol is illustrated in Fig.~\ref{fig:channel estimation}(b). The restrictions become crucial when the interactions between $\mathcal{H}$ and $\mathcal{M}$ have limited ability to change the state in $\mathcal{H}$ \footnote{When the universal gate set is available for interaction, one could use the SWAP gate between $\mathcal{H}$ and $\mathcal{M}$. The problem then becomes equivalent to conventional QCD.}. In a typical experiment, however, back-action on the channel is avoided by using a high-impedance meter. Here, we model this meter as a controlled unitary with the control on $\mathcal{H}$ and the unitary operation on $\mathcal{M}$.

\begin{figure}[t]
    \centering
    \includegraphics[width=0.95\linewidth]{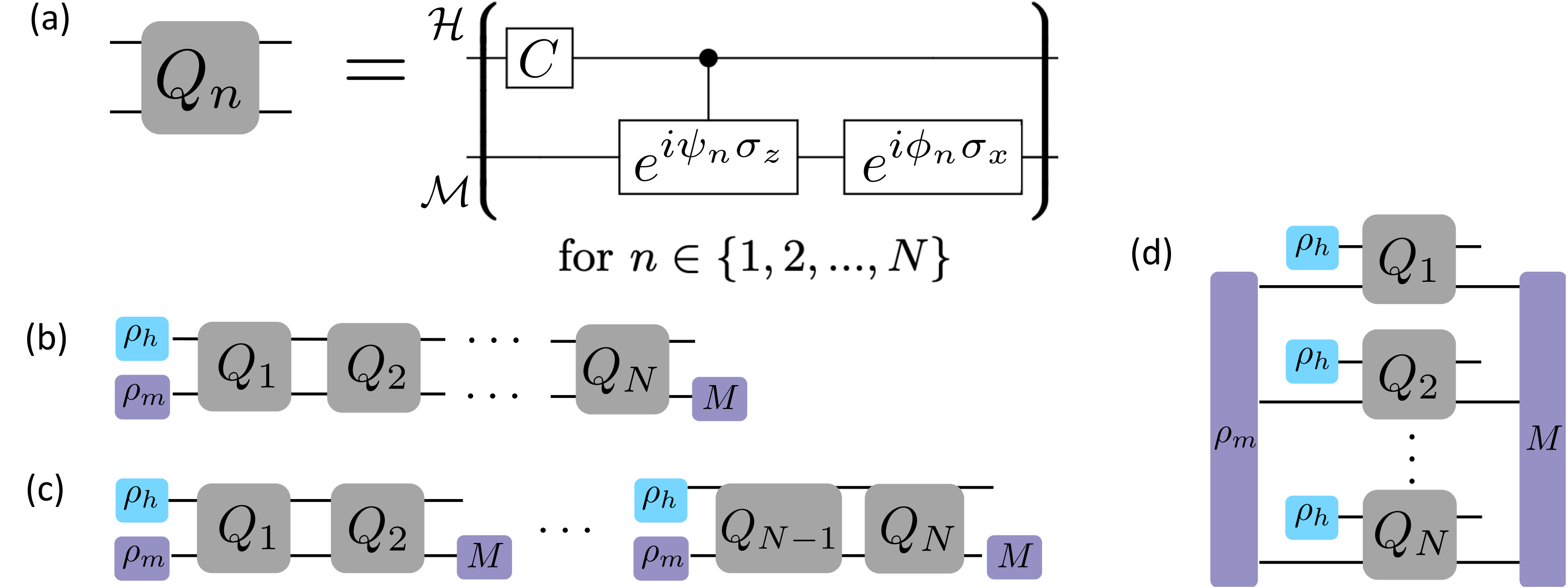}
    \caption{The query and sequential/multi-shot/parallel protocols. (a) Query $Q_n$ (Def.~\ref{def:HBCD_query}) with phases $\psi_n$ and $\phi_n$ specified independently. The upper (hidden) qubit undergoes unitary evolution every round and at the end we measure the lower (measurement) qubit.  (b-d) Discrimination protocol $S$: (b) Sequential protocol, (c) Multi-shot protocol with depth $d=2$, (d) Parallel protocol. $\rho_m$ can be a highly-entangled state for the parallel protocol.}
    \label{fig:Strategies}
\end{figure}

It is then natural to ask: 
Is discrimination with zero error probability or with fewer queries than classical methods still possible under these restrictions? This is a difficult task if conventional QCD techniques are employed. For example, discrimination of a unitary channel is impossible when the input state is the maximally-mixed state in conventional QCD \cite{helstrom}. Nevertheless, we give an affirmative answer to this question by studying Hidden Binary Channel Discrimination (HBCD), which is a minimal binary HQCD model consisting of two qubits 
as shown in Fig.~\ref{fig:Strategies}, 
and by constructing concrete measurement protocols with the desired performance. The new protocols inherit ideas from conventional QCD, including sequential, parallel, and multi-shot strategies \cite{Harrow10,adaptive_metrology_2017}, but with some surprising performance differences.

In this letter, we demonstrate that for the HBCD problem, sequential protocols outperform non-sequential protocols, including parallel and multi-shot protocols, in terms of the number of queries required to achieve a desired error probability. Furthermore, we prove that sequential protocols can achieve perfect discrimination with zero error. In contrast, we show a case where non-sequential protocols fail to solve HBCD when $C$ is applied once before measurement. We extend the quantum metrology concepts of standard quantum limit (SQL) and the Heisenberg limit to HBCD. 
The number of queries needed to solve the HBCD by sequential protocols is proven to be asymptotically optimal using an information-theoretic bound and saturates the Heisenberg limit, whereas non-sequential protocols achieve only the SQL.
These advantages of sequential protocols over parallel protocols in QCD are reported for the first time to the best of our knowledge. Finally, we illustrate how HBCD restrictions arise in an experimental example.

{\it Problem Statement: }
In our HBCD problem, we consider a two qubit system composed of a one-qubit hidden system $\mathcal{H}$ on which the unknown channel $C$ acts and a one-qubit measurement system $\mathcal{M}$ used to learn $C$. 
\begin{definition} \label{def:unknown_channel}
    {\em Unknown Channel $C$}
    Let $\alpha \in (0,2\pi)$, and   $\theta_C$ be a Bernoulli random variable taking values in $\{0,\alpha\}$ with probability $P_{\theta_C}(0)=P_{\theta_C}(\alpha)=1/2$. The unknown quantum channel acting on $\mathcal{H}$ is then $C=e^{i\theta_C \sigma_x}$.  
\end{definition}

\begin{definition} \label{def:HBCD_query}
    {\em Query.} 
    A query $Q(\psi, \phi)$ is a unitary operation that acts on the two-qubit system composed of $\mathcal{H}$ and $\mathcal{M}$, and is parametrized by a pair of phases $\{\psi, \phi\}$. The circuit of $Q(\psi, \phi)$ is depicted in Fig.~\ref{fig:Strategies}(a). It involves three components: (i) the unknown channel $C$, (ii) a controlled rotation on $\mathcal{M}$ by $\psi$ along the $z$-axis conditioned on the state of $\mathcal{H}$, and (iii) a single-qubit rotation on $\mathcal{M}$ by $\phi$ along the $x$-axis.
\end{definition}
The query as defined above is inspired from \textit{quantum signal processing} (QSP) \cite{low2017qsp} and lends to the success of the constructed protocols. Connections to QSP are elaborated in \cite{supphbcd}. We now define our HBCD problem. 
\begin{definition} \label{def:HBCD_problem}
    {\em HBCD Problem.} 
    Suppose that $\epsilon \in [0,1/2]$, and $\rho_h$ is the initial one-qubit mixed state on $\mathcal{H}$. Let $C$ be the unknown channel from Def.~\ref{def:unknown_channel} with $\theta_C$ determined at the start of the experiment and which remains constant for all subsequent queries. Then $\textsf{HBCD}(\alpha,\epsilon,\rho_h )$ defines the problem of learning an estimate $\hat{\theta}_C$ of the unknown $\theta_C$ with error probability $P(\hat{\theta}_C \neq \theta_C) \leq \epsilon$.
\end{definition}

We would ideally like to solve an HBCD problem using as few queries as possible. In addition to specifying these queries, we are allowed to specify the initial state $\rho_m$ to $\mathcal{M}$ and the POVM measurement $M$ acting on $\mathcal{M}$. Collectively, this is used to design a discrimination protocol $\Sigma$ to learn the unknown $\theta_C$. The discrimination protocols considered here involve $N$ queries $\{Q_1,\ldots,Q_{N}\}$. We denote the corresponding vector of phases as $\Phi~\equiv~(\psi_1,\ldots,\psi_N, \phi_1,\ldots, \phi_N )\in [0,2\pi)^{2N}$.

\begin{definition}
    {\rm Discrimination Protocols.} \label{def:HBCD_protocol}
    Given a problem $\textsf{HBCD}(\alpha, \epsilon, \rho_h)$, we define a discrimination protocol $\Sigma(N, d, Z, S)$ where $N$ is the total number of the queries used, depth $d$ is the number of concatenated queries before measurement, $Z=(\rho_m, \Phi, M)$ is the collection of specified settings with $\Phi$ being the vector of phases specifying the $N$ queries, and $S$ defines the type of protocol which can be sequential, multi-shot or parallel. The circuit corresponding to $\Sigma$ for different $S$ is shown in Fig.~\ref{fig:Strategies}(b)-(d). 
\end{definition}
Note that our discrimination protocols are designed using knowledge of $\rho_h$ and $\alpha$. The depth $d$ takes the value of $N$ when $S$ is sequential, $N/m$ when $S$ is a multi-shot protocol using $m$ shots and $1$ when $S$ is a parallel protocol over an $N$-qubit measurement system $\mathcal{M}$ interacting with $N$ copies of $\mathcal{H}$ (see Fig.~\ref{fig:Strategies}(d)). 
Our sequential protocol uses one probe qubit \footnote{We note that if the sequential protocol can operate on $N$ qubits and arbitrary operations can be used, it is strictly stronger than parallel protocols \cite{Yuan16, Bararesco21}.}, which has weaker discrimination performance compared to the parallel protocol in conventional QCD \cite{Piani09,BaePiani19}. We compare their performances in HBCD. The multi-shot protocol allows for adaptive choice of $Z$, but we do not explore it in this letter \footnote{When $d$ is fixed, we do not expect adaptivity to change the asymptotic scaling of $N$ with $\alpha$ as observed in the case of conventional QCD \cite{Cooney16,salek_adaptive_2021}. When $d$ is allowed to adaptively change, a higher scaling may be achieved. However, this is already captured by the sequential protocol.}.

Let us now define the discrimination error associated with each protocol. Suppose $(y^1,\ldots,y^m)$ is a set of $m$ POVM outcomes, collectively denoted by the vector $\mathbf{y}\in\{0,1\}^m$. Given $\mathbf{y}$, an estimator $\hat{\theta}_C(\mathbf{y})$ will output either $0$ or $\alpha$. The error probability of a protocol $\Sigma$ is then
\begin{align}
    \nonumber &P(\hat{\theta}_C(\mathbf{y}) \neq \theta_C ; \Sigma) \\ 
    &= \frac{1}{2}\left[P_{\hat{\theta}_C(\mathbf{y})|\theta_C}(\alpha| 0 ; \Sigma) + P_{\hat{\theta}_C(\mathbf{y})|\theta_C}( 0| \alpha ; \Sigma)\right],
    \label{eq:estimation_error_hbcd}
\end{align}
where we have used the fact that the prior probabilities satisfy $P(\theta_C = 0) = P(\theta_C = \alpha)=1/2$. The estimator is designed such that $\hat{\theta}_C(0)=0$ and $\hat{\theta}_C(1)=\alpha$. Therefore, the error is given by $\frac{1}{2}(P_{y|\theta_C}(0|\alpha) + P_{y|\theta_C}(1|0))$. For the multi-shot and parallel protocols, we use the likelihood ratio test as our estimator. An overview of estimators is given in \cite{supphbcd}.

Another metric of performance of the protocols is the scaling of the minimal number of queries $N$ required to solve $\textsf{HBCD}(\alpha, \epsilon, \rho_h)$. As $\alpha$ becomes smaller, solving HBCD becomes more difficult and hence $N$ should grow. We can then define two scaling limits of $N$ with $\alpha$.
\begin{definition}
    {\rm Standard quantum limit and Heisenberg scaling in HBCD.} \label{def:HBCD_scaling}
    Suppose that $\epsilon \in [0,1/2)$, $\rho_h$, $S$, and $d$ are given. For $0 < \alpha \ll 1$, let $N(\alpha)$ be the number of queries needed to solve $\textsf{HBCD}(\alpha, \epsilon, \rho_h)$ by $\Sigma(N,d,Z,S)$. We say that a depth-$d$ $S$ protocol achieves the standard quantum limit (SQL) if $N(\alpha)=\Theta(\alpha^{-2})$ and Heisenberg scaling if $N=\Theta(\alpha^{-1})$.
\end{definition}
The SQL and the Heisenberg limit are defined in quantum metrology for parameter estimation in terms of the number of access to an unknown physical system of interest. This corresponds to the number of interactions $N$ between $\mathcal{H}$ and $\mathcal{M}$. In parameter estimation, a protocol is said to achieve SQL when the number of queries $N$ required to achieve an estimation error $\alpha_{\rm PE}$ scales as $N\sim \alpha_{\rm PE}^{-2}$ and the Heisenberg limit when $N\sim \alpha_{\rm PE}^{-1}$ \cite{giovannetti_2011, Braun18, Pezze18}. Similarly, we can model the problem of discriminating the value of $\theta_C$ from $\{0,\alpha\}$ in HBCD as estimating the value of $\theta_C$. In this case, we succeed if the estimation error is smaller than half of the angle difference $(\alpha/2)$. Definition~\ref{def:HBCD_scaling} is then evident.

{\it Perfect discrimination: }
We now discuss advantages of using sequential protocols in HBCD. The proofs of the theorems presented below are in \cite{supphbcd}.

\begin{theorem}\label{thrm: perfect discrimination}
    {\em Perfect discrimination in HBCD.}

    For all $\alpha \in (0,2\pi)$, there exists a sequential protocol $\Sigma(N, d=N, Z, S=sequential)$ that solves HBCD$(\alpha, \epsilon=0, \rho_h)$ with at most $N=j\lceil \frac{2\pi}{\beta} \rceil$ queries. 
\end{theorem} 
Here $j$ is 1 for $\alpha \in [0,\frac{\pi}{4}] \cup [\frac{3\pi}{4},\frac{5\pi}{4}] \cup [\frac{7\pi}{4},2\pi)$, 2 for $[\frac{3\pi}{8},\frac{5\pi}{8}] \cup [\frac{11\pi}{8},\frac{13\pi}{8}]$ and 3 for the rest. Next $\beta$ is an effective rotation angle on the measurement qubit (shown in \cite{supphbcd}). To prove the theorem, we first make a diagonal unitary matrix with four query iterations. The controlled rotation then becomes a single-qubit $R_Z$ gate on $\mathcal{M}$ with its rotation angle being either $-\beta$ for $\theta_C=0$ or $\beta$ for $\theta_C=\alpha$. Using this rotation on $\mathcal{M}$, we accumulate the phase $\pm \beta$ in the measurement qubit so that measurement qubit is $\state{0}$ for $\theta_C=0$ and $\state{1}$ for $\theta_C=\alpha$ \cite{rossi2021quantum, mrtc_2021_unification}. Although the above sequential protocol only achieves the SQL, our numerical results and information-theoretic bound show that sequential protocols can be designed to attain the Heisenberg limit.

{\it Weakness of the non-sequential protocol:}
Since any entanglement improves conventional QCD \cite{Piani09, BaePiani19}, one may expect parallel protocols to be better than sequential protocols. Indeed, this expectation is valid when the channel is noisy and error correction is unavailable \cite{zhou_2018}. 
However, when the process is noiseless but the state is noisy, the opposite is true;  
the sequential protocol outperforms the parallel protocol. 
When the query depth $d=1$ and the initial state in $\mathcal{H}$ is maximally mixed, we show that discrimination is impossible for non-sequential protocols. 
\begin{theorem}\label{thrm: impossibility by parallel}
    {\em Impossible case for depth-1 non-sequential protocols}.
    Suppose that $\rho_h = \frac{\mathbb{I}}{2}$, and $S$ is the multi-shot or parallel protocol. For any $\epsilon \in [0,1/2)$ and $\alpha \in (0,2\pi)$, the protocols $\Sigma(N, d=1,Z, S)$ cannot solve $\textsf{HBCD}(\alpha,\epsilon, \rho_h)$. That is, $\Sigma$ does not obtain any information on $\theta_C$ through $M$.
\end{theorem} 

The key idea to the proof is that the maximally mixed state remains invariant under single qubit rotations. 
Therefore, the state ($\rho_M$) of $\mathcal{M}$ before measurement does not depend on the value of $\theta_C$. However, if $d\geq 2$ queries are used, $\rho_M$ correlates with $\theta_C$ through the controlled interaction.
Thus, protocols with $d\geq 2$ queries are strictly better than non-sequential protocols of $d=1$. Next, we prove the asymptotic number of queries required to solve the HBCD for $d = 2$. The multi-shot protocol with a fixed depth cannot achieve the Heisenberg limit (Theorem~\ref{thrm: SQL}), which is illustrated numerically later.
\begin{theorem}\label{thrm: SQL}
    {\em Standard quantum limit in HBCD by multi-shot protocol}. 
    For all $\epsilon \in [0,1/2)$ and $\rho_h$, depth-$2$ multi-shot protocol achieves the SQL. 
\end{theorem} 
The theorem implies that HBCD becomes challenging with decreasing $\alpha$, and the minimum distinguishable value of $\alpha$ scales as $\alpha \sim N^{-1/2}$ with increasing $N$.

The advantages of sequential protocols over non-sequential protocols in HBCD are evident from Theorems~\ref{thrm: perfect discrimination}-\ref{thrm: SQL}. The sequential protocol alone enables perfect discrimination. Additionally, non-sequential protocols with $d=1$ cannot determine $\theta_C$ regardless of the number of queries, while the sequential protocol can.

{\it Heisenberg limit in HBCD: }
The possibility of achieving the Heisenberg limit (Def.~\ref{def:HBCD_scaling}) is still unanswered. We first derive a lower bound on $N$ required to solve HBCD.
\begin{theorem}\label{thrm:lower_bound}
    {\em Fundamental limit of HBCD}. 
    Any protocol $\Sigma(N,d,Z,S)$ with $N < \frac{1}{\sqrt{2(1- \cos \alpha)}}$ cannot solve $\textsf{HBCD}(\alpha,\epsilon=0,\rho_h)$.
\end{theorem} 
Expanding the bound on $N$ in Theorem~\ref{thrm:lower_bound} around $\alpha \ll 1$ gives us that the Heisenberg limit is indeed the optimal scaling, i.e., $N=\Omega(\alpha^{-1})$.

\begin{figure}[h!]
    \centering
    \includegraphics[width=0.95\linewidth]{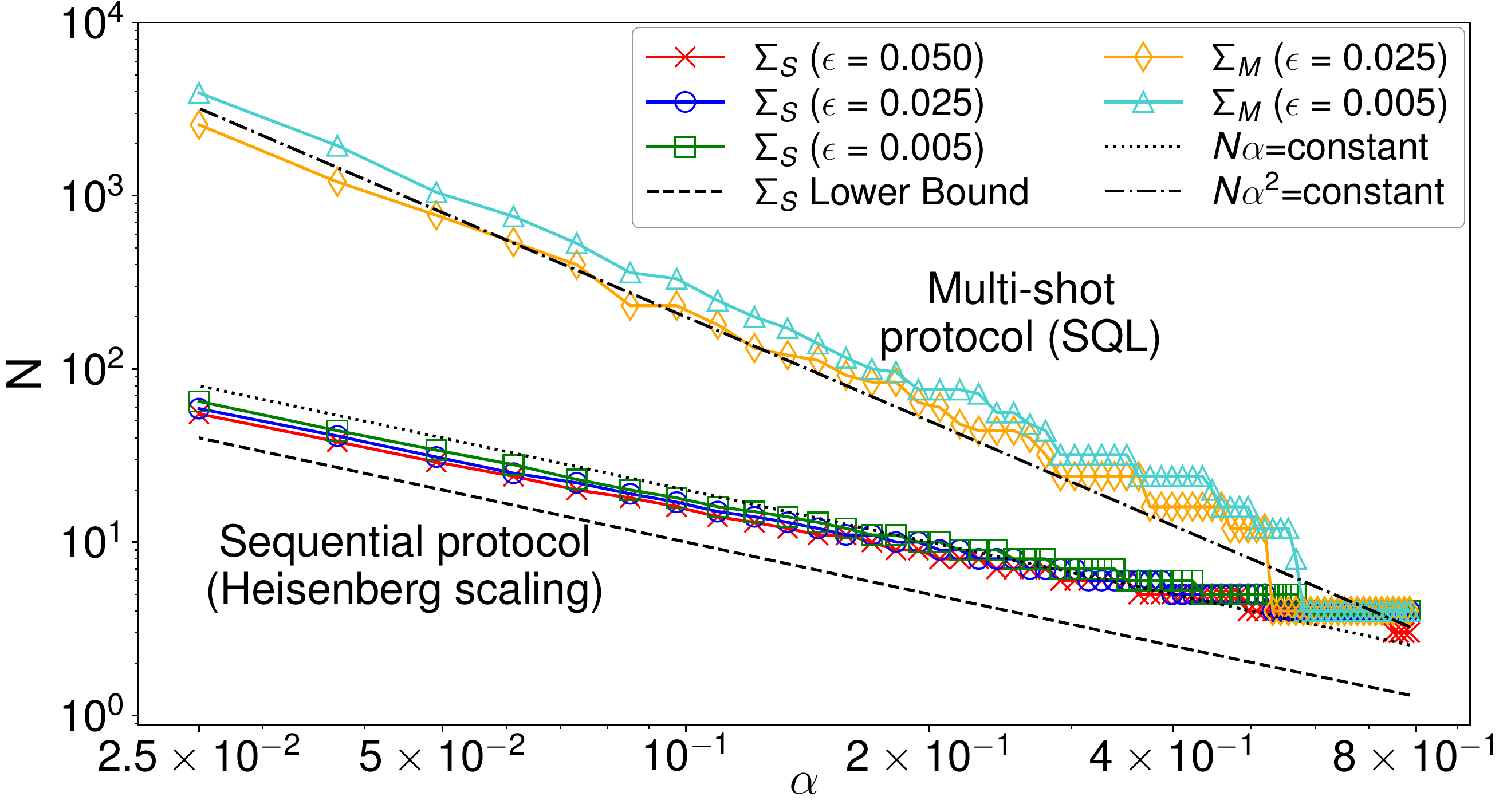}
    \caption{Number of queries $N$ sufficient for solving $\textsf{HBCD}(\alpha, \epsilon, \rho_h = \mathbb{I}/2)$. For the sequential protocol ($\Sigma_S$), we measure only once and use a phase sequence $\Phi$ of length $N$. For the multi-shot protocol ($\Sigma_M$), we use queries of depth $d=4$ and measure multiple times. Trends for different values of $\epsilon \in \{0.005, 0.025, 0.05\}$ are shown for $\Sigma_S$ and $\epsilon \in \{0.025, 0.005\}$ for $\Sigma_M$.}
    \label{fig:numerical_bounds_hbcd}
\end{figure}

We now present numerical evidence that the Heisenberg limit is indeed achieved by the sequential protocol while the multi-shot protocol using constant depth queries only achieves the SQL. We solve the HBCD problem through measurements on $\mathcal{M}$ shown in Figure~\ref{fig:Strategies}(b,c) using a maximally mixed state ($\rho_h=\frac{\mathbb I}{2}$) on $\mathcal{H}$. The state on $\mathcal{M}$ depends on the specified phase sequence $\Phi$. If some $\Phi$ of length $N$ sets the state of $\mathcal{M}$ to be $|1\rangle$ for $\theta_C = \alpha$ and $|0\rangle$ for $\theta_C = 0$, then $\textsf{HBCD}(\alpha,\epsilon=0,\rho_h)$ is solved with one shot.

For the sequential protocol, we attempt to solve HBCD by measuring once and with error probability $\epsilon \in [0,1/2)$. The goal is to set the outcome $y$ of measuring $\mathcal{M}$ in the computational basis such that
\begin{equation}
    P_{y|\theta_C}(1|\alpha) - P_{y|\theta_C}(1|0) \geq 1 - 2 \epsilon,
    \label{eq:prob_measurements_hidden_bcd}
\end{equation}
where we have used Eq.~\ref{eq:estimation_error_hbcd}. To determine $\Phi$ that satisfies Eq.~\ref{eq:prob_measurements_hidden_bcd}, we solve the following optimization problem
\begin{equation}
    \arg \min_{\Phi} \left(1 - P_{y|\theta_C}(1|\alpha;\Phi) + P_{y|\theta_C}(1|0;\Phi) \right)^2,
    \label{eq:optimization_prob_hidden_BCD}
\end{equation}
with the additional constraint $\psi_n=\psi, \forall n \in [N]$. Details of the optimization is given in \cite{supphbcd}. We claim that $\Phi$ succeeds in $\textsf{HBCD}(\alpha, \epsilon, \rho_h)$ if the solution to Eq.~(\ref{eq:optimization_prob_hidden_BCD}) satisfies Eq.~(\ref{eq:prob_measurements_hidden_bcd}), i.e., $R(\Phi) \leq 4\epsilon^2$ where $R(\cdot)$ corresponds to the loss function defined inside Eq.~\ref{eq:optimization_prob_hidden_BCD}. Given $\alpha$, we determine the minimal number of queries required by starting with $N=1$ and incrementing the value of $N$ by one until the solution to Eq.~\ref{eq:optimization_prob_hidden_BCD} satisfies Eq.~\ref{eq:prob_measurements_hidden_bcd}. 

For the multi-shot protocol with constant depth-$d$ queries, we use a phase sequence $\Phi$ of length $d$ but may measure $\mathcal{M}$ $m\geq 1$ times to solve HBCD with error probability $\epsilon \in [0,1/2)$. For a given value of $\alpha$, we determine $\Phi$ of length $d$ by solving the optimization problem of Eq.~\ref{eq:optimization_prob_hidden_BCD}. We determine $m^\star$ or the smallest number of shots required to achieve an error $\epsilon$ by evaluating Eq.~\ref{eq:estimation_error_hbcd}, considering the estimator based on the likelihood-ratio test over the measurement outcomes \cite{supphbcd}. The total number of queries required is then $N=d \cdot m^\star$.

In Fig.~\ref{fig:numerical_bounds_hbcd}, we show the numerically determined trends of $N$ required by the sequential and multi-shot protocols to solve $\textsf{HBCD}(\alpha, \epsilon, \rho_h=\mathbb{I}/2)$. As expected, $N$ increases as $\alpha$ decreases and approaches zero for both protocols. In particular, we observe a scaling of SQL for the multi-shot protocol but crucially a Heisenberg limited scaling $N \sim O(\alpha^{-1})$ for the sequential protocol.

\begin{figure}[t]
    \centering
    \includegraphics[width=0.48 \textwidth]{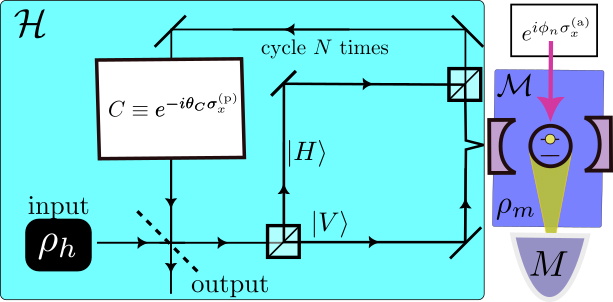}
    \caption{ The schematic for optical experiment for measurements through hidden channel discrimination. The color code is identical to that in Fig.~\ref{fig:channel estimation}(a) and Fig.~\ref{fig:Strategies}(a), with the photonic (p) and atomic (a) degrees of freedom realizing systems $\mathcal{H}$ and $\mathcal{M}$, respectively. See~\cite{supphbcd} for details of implementation.
    }
    \label{fig:Exp}
\end{figure}

{\it HBCD Example- }
To demonstrate the advantage of the sequential protocol in HBCD in a realistic setting, consider the discrimination of an unknown channel 
describing the presence or absence of a birefringent slab which rotates the polarization of incident single photons by an angle $\alpha$. 
Naively, the discrimination of a birefringent slab can be formulated as a conventional QCD in Fig.~\ref{fig:channel estimation}(a) by using propagating a polarization qubit. Sequential protocols are known to solve conventional QCD efficiently \cite{Braun18, rossi2021quantum,elitzur1993quantum}.

However, in the free-space setting, optical elements that implement $\{V_n\}$ cannot be reconfigured at the timescale of a roundtrip \cite{saleh2019fundamentals,barz2015quantum}. Consequently, a sequential protocol without an adjoint measurement system $\mathcal{M}$ requires the number of optical elements to increase with the depth of the protocol. 
In other words, the propagating photonic polarization qubit should be considered a hidden system $\mathcal{H}$ for small enough $\alpha$. 

To address this problem, we introduce a measurement system $\mathcal{M}$ that consists of a cavity QED system (see Fig.~\ref{fig:Exp}). $\mathcal{M}$ not only enables the implementation of reconfigurable single-qubit gates, but also provides improved measurement efficiency and fidelity 
\cite{burrell2010high,supphbcd}.
The interaction between $\mathcal{H}$ and $\mathcal{M}$ required for the implementation of the query $Q$ is realized by the protocol proposed in \cite{duan2004scalable}.
With this implementation, Theorem~\ref{thrm: perfect discrimination} and the numerical result in Fig.~\ref{fig:numerical_bounds_hbcd} provides a concrete protocol for perfect discrimination and Heisenberg scaling in discriminating the presence or absence of a birefringent slab.

Although decoherence may prevent the Heisenberg limit from being saturated in practice, Fig.~\ref{fig:numerical_bounds_hbcd} indicates that the proposed HBCD sequential protocol still offers advantages in achieving higher probability of detection and lower error even for shallower protocols with weaker requirements on query fidelity \cite{supphbcd}. 

{\it Conclusion: }
In this letter, we investigated HBCD with different protocols and analyzed their performance. We showed that the sequential protocol has concrete advantages in HBCD over non-sequential protocols: (i) it achieves perfect discrimination whereas the parallel and multi-shot protocols with query depth $d=1$ cannot be used to learn the unknown channel, and (ii) it saturates the Heisenberg limit while the multi-shot protocol with $d\geq 2$ can only solve HBCD at the SQL. Further, we demonstrate the sequential protocol can be naturally realized in an experimental setup based on cavity QED. 

{\it Code and data availability: } Code for the different discrimination protocols in solving HBCD numerically and data are available on GitHub~\footnote{\url{https://github.com/arkopaldutt/HiddenBCD}}.

{\it Acknowledgments}
The authors thank Zane M. Rossi, Victor M. Bastidas, and Yoshihisa Yamamoto for useful discussions and valuable comments.  AD and ILC were supported in part by the U.S. Department of Energy, Office of Science, National Quantum Information Science Research Centers, and Co-Design Center for Quantum Advantage under contract DE-SC0012704.  We also acknowledge NTT Research for its financial and technical support. 

\bibliography{references}

\clearpage

\onecolumngrid
\appendix

{\begin{center}\textbf{\Large Supplementary Material: Power of sequential protocols in hidden quantum channel discrimination}\end{center}
\vspace{0.25in}}

In Section~\ref{app_sec:QSP}, we give motivation for the query used in our discrimination protocols for solving the HBCD problem. In Section~\ref{app_sec:estimators}, we give an overview of estimators applicable to our protocols. In Sections~\ref{app: thrm1}-\ref{app_sec:thrm4}, we give proofs of the theorems stated in the main paper which comment on the performance of the different protocols investigated in this work and the minimal number of queries required for a sequential protocol to solve a given HBCD problem. In Section~\ref{app_sec:numerical_experiments}, we give details on the numerical experiments on assessing the performance of the sequential and multi-shot protocols on different HBCD problems. In Section~\ref{app_sec:operating_characteristics_hbcd}, we describe how operating characteristics for different HBCD protocols can be defined. Further, we assess the performance of the sequential and multi-shot protocols in terms of probability of detection under different constraints. Finally in Section~\ref{app_sec:HBCD_example}, we give details of the experimental example proposed in the paper. We discuss the experimental restrictions on measuring and preparing single-photons. Moreover, we calculate the error bound on the detection probability using the subadditivity of errors. This bound allows us to evaluate the performance of the proposed experimental implementation in the presence of experimentally relevant decoherence mechanisms. Finally we compare our protocol to those that use Gaussian photonic states. 

\section{Relation to Quantum Signal Processing}\label{app_sec:QSP}
The query (Def.~\ref{def:HBCD_query}) as used in our discrimination protocols to solve HBCD is inspired from Quantum Signal Processing (QSP). The qubit in the measurement system $\mathcal{M}$ (see Fig.~\ref{fig:Strategies}) can be viewed as the ancilla qubit typically used in QSP \cite{low2017qsp} with the rotations of $\exp(i \phi_n \sigma_x)$ for any $n \in [N]$ as the processing operators. The signal operators then involve the application $C$ on $\mathcal{H}$, and the controlled interaction on the composed system of $\mathcal{H}$ and $\mathcal{M}$. Note that the signal operator in this case looks different from that typically used in QSP. 

Moreover in connection to QSP, tracing out the hidden system $\mathcal{H}$ of the query results in a Completely Positive Trace-Preserving map with non-Markovian process, which can be interpreted as a noisy channel. In the sequential protocol, the query involves the application of a tunable X-rotation gate and a noisy channel alternately, which has a structure of QSP. However, it is not expected that the number of the query used to solve the HBCD saturates the Heisenberg limit with a noisy channel. 

The result changes when partial information in $\mathcal{H}$ is known. In particular, it is noted that although the initial state and rotation angle of $C$ are still unknown, it is known that $C$ is the X-rotation gate and that the interaction between $\mathcal{H}$ and $\mathcal{M}$ is a controlled rotation. In this scenario, the result shows that the Heisenberg limit and perfect discrimination can be achieved using Quantum Signal Processing (QSP). This is an interesting finding as it suggests that even with partial information, QSP can be utilized to achieve highly precise measurements and accurate discrimination.

\section{Estimators}\label{app_sec:estimators}
In this section, we describe the different estimators that can be used for solving HBCD with different protocols. Let the measurement outcomes from applying any protocol be given by $y^k \in \{0,1\}$, indexed by $k$. We collectively denote the vector of $m$ binary outcomes as $\mathbf{y} = (y^1,\ldots,y^m) \in \{0,1\}^m$. Suppose we set the phases $\Phi$ corresponding to the protocol such that $y=1$ with a high probability for $\theta_C = \alpha$ and $y=0$ with a high probability for $\theta_C = 0$. Some estimators that can then be used are as follows.

\paragraph{Majority Vote:} A simple (albeit suboptimal) estimator for $\hat{\theta}_C$ uses the majority vote (denoted by $\mathrm{Maj}$) of the measurement outcomes:
\begin{equation}
    \hat{\theta}_C = \alpha \cdot \mathrm{Maj}(\mathbf{y}).
\end{equation}

\paragraph{Likelihood Ratio Test:} The likelihood ratio test (LRT) or the maximum likelihood estimator is the optimal estimator for binary hypothesis testing.

Consider the log-likelihood function:
\begin{align}
    L(\mathbf{y};\theta_C) &= \sum_{k=1}^m \log p(y^k | \theta_C) \\
    &= Y_1 \log p(1 | \theta_C) + Y_0 \log p(0 | \theta_C),
\end{align}
where $Y_1 = \sum_k y^k$ and $Y_0 = m - Y_1$

From maximum likelihood, we then have that
\begin{equation}
    \hat{\theta}_C = 
    \begin{cases}
    \alpha &, \, L(\mathbf{y};\alpha) > L(\mathbf{y}; 0) \\
    0 &, \, L(\mathbf{y};\alpha) \leq L(\mathbf{y}; 0)
    \end{cases}
    \label{eq:lrt_hbcd}
\end{equation}

\section{Proof of Theorem \ref{thrm: perfect discrimination}}\label{app: thrm1}
First, we assume $\alpha \in  \mathcal{D}_1 =[0,\frac{\pi}{4}] \cup [\frac{3\pi}{4},\frac{5\pi}{4}] \cup [\frac{7\pi}{4},2\pi)$ and prove that the perfect discrimination is possible in this case. Later we extend it for all $\alpha$. Let us consider a unitary matrix 
\begin{align}
    \check{Q}(\theta_C, \psi)=Q_4 Q_3 Q_2 Q_1 
\end{align}
with $\phi_n=0$ and $\psi_n=\psi$ for $\forall n$. In this section we use computational basis $\{\state{00},\state{01},\state{10},\state{11}\}$ where the first qubit is in $\mathcal{H}$ and the second qubit is in $\mathcal{M}$. Then the matrix elements of $\check{Q}$ is given as follows: 
\begin{align}
    \check{Q}=
    \begin{bmatrix}
        P_1(x, a) && 0 && iR(x,a) &&0\\
        0&& P_2(x,a) && 0 && i{\frac{1}{a^k}} R(x,a) \\
        iR(x,a) && 0 && P_3(x,a) &&0\\
        0&& i{1\over a^{k-1}}R(x,a) && 0 && P_4(x,a) ,
    \end{bmatrix}
    \label{eq: U_simple 4}
\end{align} 
where we use parametrizations $x\equiv \cos \theta_C$ and $a\equiv \exp(i \psi)$, $P_i(x,a)$ is a 4-degree polynomial of $x$ and $a$, which is even in $\theta_C$, 
\begin{align}
    P_1(x,a)&=a^2 - a (1 + a) (3 + a) x^2 + (1 + a)^3 x^4\\
    P_2(x,a)&=\frac{a - (1 + a) (1 + 3 a) x^2 + (1 + a)^3 x^4}{a^3}\\
    P_3(x,a)&=a (a - (1 + a) (1 + 3 a) x^2 + (1 + a)^3 x^4)\\
    P_4(x,a)&=\frac{a^2 - a (1 + a) (3 + a) x^2 + (1 + a)^3 x^4}{a^4},
\end{align}
and $R(x,a)$ is a function that has a following form 
\begin{align}
    R(x,a)=x\sqrt{1-x^2}a(1-a)(-2 a + (1 + a)^2 x^2). 
\end{align}

An important observation for $\check{Q}$ is that all the off-diagonal elements share $R(x,a)$. 
For a given $x$ there exists $\tilde{a}(x)$ such that
\begin{align}
    R(x,\tilde{a}(x))=0, \label{eq: condition diagonal}
\end{align}
if and only if $\theta_C \in  \mathcal{D}_1$.  

Now, we know that $\theta_C$ is a Bernoulli random variable in our HBCD, taking a value of $0$ or $\alpha$. Since we assume that $\alpha\in \mathcal{D}_1$, we choose $a=\tilde{a}(\alpha)$. Then we can readily show that $\check{Q}$ becomes diagonal both for $\theta_C=0$ and for $\theta_C=\alpha$: 
\begin{align}
    \check{Q}\left(0, -i \log (\tilde{a}) 
    \right)=
    \begin{bmatrix}
        1 && 0 && 0 &&0\\
        0&& 1 && 0 && 0 \\
        0 && 0 && e^{2i \beta} &&0\\
        0&& 0 && 0 && e^{-2i \beta} 
    \end{bmatrix}, \label{eq: Usimple general 0}
\end{align}
and 
\begin{align}
    \check{Q}\left(\alpha, -i \log (\tilde{a}) 
    \right)=
    \begin{bmatrix}
        e^{i \beta} && 0 && 0 &&0\\
        0&& e^{-i \beta} && 0 && 0 \\
        0 && 0 && e^{i \beta} &&0\\
        0&& 0 && 0 && e^{-i \beta}.
    \end{bmatrix},
    \label{eq: Usimple general nonzero}
\end{align}
where the rotation angle $\beta$ is given by $\beta = -i\log (P_1(\cos{\alpha}, \tilde{a}))$. 

Now we can obtain an analytical solution that achieves the perfect discrimination between $\theta_C=0$ and $\theta_C=\alpha$ when $\alpha\in \mathcal{D}$. First, when $\beta$ has the form of $\beta = {\pi \over 2 n}$ ($n\in \mathbb{N}$), we make a $n$-th power of $\check{Q}^n$. 
\begin{align}
    \check{Q}^n\left(\theta_C=0, \psi= -i \log (a_{\rm sol}) \right)=
    \begin{bmatrix}
        1 && 0 && 0 &&0\\
        0&& 1 && 0 && 0 \\
        0 && 0 && -1 &&0\\
        0&& 0 && 0 && -1 
    \end{bmatrix}, \label{eq: Usimple nth 1}
\end{align}
and 
\begin{align}
    \check{Q}^n\left(\theta_C=\alpha, \psi= -i \log (a_{\rm sol})\right)=
    \begin{bmatrix}
        i && 0 && 0 &&0\\
        0&& -i && 0 && 0 \\
        0 && 0 && i &&0\\
        0&& 0 && 0 && -i.
    \end{bmatrix},
    \label{eq: Usimple nth 2}
\end{align}

In this case, perfect discrimination is done with $N=4n$ queries with $\rho_m =\frac{1}{2}(\state{0}+i \state{1})(\bstate{0}-i\bstate{1})$ and $\vec{\phi}_{\rm sol}\equiv \{\phi_1,\cdots \phi_{4n} \}= \{ 0,0,\cdots,0 , {\pi\over 4}\}$. 
The overall unitary matrix $U=Q_{4N} \cdots Q_1$ reads:   
\begin{align}
    U\left(\theta_C=0
    \right)=
    \frac{1}{\sqrt{2}}
    \begin{bmatrix}
        1 && i && 0 &&0\\
        i&& 1 && 0 && 0 \\
        0 && 0 && -1 && -i\\
        0&& 0 && -i &&  -1    
    \end{bmatrix}, \label{eq: Usol zero}
\end{align}
and 
\begin{align}
    U\left(\theta_C=\alpha
    \right)=
    \frac{1}{\sqrt{2}}
    \begin{bmatrix}
        i && 1 && 0 &&0\\
        -1&& -i && 0 && 0 \\
        0 && 0 && i && 1\\
        0&& 0 && -1 && -i  
    \end{bmatrix}. \label{eq: Usol nonzero}
\end{align}
We can readily check that regardless the state in $\mathcal{H}$ the final state in $\mathcal{M}$ is $\state{1}$ for $\theta_C=0$ and $\state{0}$ for $\theta_C=\alpha$. Therefore this unitary achieves perfect discrimination. When ${\pi \over 2 (n-1)} <\beta <{\pi \over 2 n}$, we can also construct the function that satisfies Eqs.~\eqref{eq: Usol zero} and \eqref{eq: Usol nonzero} by choosing the angles $\phi_n$ \cite{Rossi22}. Therefore proof is done for $\alpha\in \mathcal{D}_1$. 

When $\alpha\notin \mathcal{D}_1$, we consider a process 
\begin{align}
    S_2=Q_2 Q_1 
\end{align}    
for $\alpha \in \mathcal{D}_2 = [\frac{3\pi}{8},\frac{5\pi}{8}] \cup [\frac{11\pi}{8},\frac{13\pi}{8}]$ and 
\begin{align}
    S_3=Q_3 Q_2 Q_1 
\end{align}    
for the rest, i.e. $\alpha \in \mathcal{D}_3 = [\frac{\pi}{4},\frac{3\pi}{8}] \cup 
[\frac{5\pi}{8},\frac{3\pi}{4}] \cup
[\frac{5\pi}{4},\frac{11\pi}{8}] \cup
[\frac{13\pi}
{8},\frac{7\pi}{4}]$ . In both cases, we choose $\psi_n=\phi_n=0$ for all $n$. Then we show that $S$ has a form
\begin{align}
    S_n=
    \begin{bmatrix}
        \cos(n\theta_C) && i \sin(n\theta_C) && 0 &&0\\
        i \sin(n\theta_C)&&  \cos(n\theta_C) && 0 && 0 \\
        0 && 0 &&         \cos(n\theta_C) && i \sin(n\theta_C) \\
        0&& 0 && i \sin(n\theta_C) &&  \cos(n\theta_C)
    \end{bmatrix}, \label{eq: Sn general}
\end{align}
which is nothing but $R_x(n\theta_C)$ rotation gate on $\mathcal{H}$. We can readily show that $2\alpha \in \mathcal{D}_1$ for $\alpha \in \mathcal{D}_2$, and $3\alpha \in \mathcal{D}_1$ for $\alpha \in \mathcal{D}_3$. Therefore both cases reduce to the first case, $\alpha \in \mathcal{D}_1$.

An important consequence of the fact that the polynomials $P_i$ are even polynomials of the unknown angle $\theta_C$ is that the discrimination protocol does not depend on the initial state of the hidden system. First consider the situation that the initial state of the hidden system is pure. Then it can be expressed in the eigenbasis of the channel Hamiltonian (i.e., $\sigma_{x}$) 
\begin{align}
\ket{\psi_{\mathcal{H}}} = \sqrt{\alpha } \ket{+} + \sqrt{1-\alpha}\ket{-}.
\end{align}
Applying the proposed composite protocol $U(\theta_C= 0)$ to this initial state then $\mathcal{M}$ does not flip independently of $\ket{\psi_{\mathcal{H}}}$. Similarly, applying $U(\theta_C= 0)$ to the composite systems, we see $\mathcal{M}$ is flipped independently of $\ket{\phi_{\mathcal{H}}}$. As a result, when $\alpha \in  \mathcal{D}$, then the discrimination protocol is independent of the initial state.

Lastly, we briefly mention why this solution only achieves the SQL. The number of query needs for this solution is characterized by $\beta$. Let us assume that $\alpha$ is small. Then $x=\cos (\alpha) = 1- \frac{\alpha}{2} + O)(\alpha^2)$. By solving $R(x,a)=0$ and $\beta = -i\log (P_1(\cos{\alpha}, \tilde{a}))$ for small $\alpha$, we obtain $\beta = 2\alpha^2$. This means that the solution in Theorem \ref{thrm: perfect discrimination} obeys the SQL. 

\section{Proof of Theorem \ref{thrm: impossibility by parallel}}
We show it by calculating the quantum state in the circuit step by step. After applying $C$, the state becomes $\rho_h \otimes \rho_m$. 
Notice that the rotation on the hidden qubit does not change $\rho_h$ because it is the maximally mixed state. Then we apply the controlled rotation gate and $e^{i \phi_1 \sigma_z}$, and obtain 
\begin{align}
    \frac{1}{2}(\ket{0}\bra{0}R_x(\phi_1)R_z(\psi_1)\rho_mR_z(-\psi_1)R_x(-\phi_1)+\ket{1}\bra{1}R_x(\phi_1)\rho_m R_x(-\phi_1)).
\end{align} 
This is independent of $\theta_C$, and thus, the measurement outcome of the measurement qubit does not determine $\theta_C$ at all. 
The situation does not change even when the parallel protocol is used. Since $R_x(\theta_C)^{\otimes N} 1^{\otimes N} R_x^\dagger(\theta_C)^{\otimes N}= 1^{\otimes N}$, the rotations on the hidden qubits do not change the state of the hidden qubits through the controlled rotations. Thus, the measurement qubit does not depend on whether $\theta_C=0$ ore $\alpha$. $\blacksquare$ 

\section{Proof of Theorem \ref{thrm: SQL}}\label{app_sec:thrm3}
Here we compute the asymptotic scaling of a multi-shot protocol with constant query depth, $d={\rm const}$. Since the asymptotic scaling does not change for adaptive protocols \cite{Cooney16}, we consider non-adaptive protocols. Suppose the minimum error probability of a single shot is $p_{\rm s}$. Assume that the minimization is done for $\rho_i$, $\phi_i$,$\psi_n$, and $M$, we estimate the lower bound for $p_{\rm s}$. 

The minimum error probability to distinguish two pure states is given by the Helstrom bound. In distinguishing two quantum channels, the minimum error probability of single-shot measurement $P_{\rm s}$ is obtained by minimizing the Helstrom bound over possible input states. 

Suppose that we have a quantum circuit $U$ to discriminate two channels. The action of $U$ is different for $\theta_C=0$ and for $\theta_C=\alpha$. Let $U_i$ $(i=1,2)$ be a unitary operator in the estimation protocol before measurement for $\theta_C=0$ and $\alpha$, respectively. Then the error probability of a one-shot measurement is given as follows. 
\begin{align}
    P_{\rm s}(\hat{\theta}_C \neq \theta_C)= \min_{\state{\psi}} \frac{1-\sqrt{1-|\bstate{\psi}U_1^\dagger U_2\state{\psi}|^2}}{2}, 
\end{align}
where $\state{\psi}$ is the initial state. Since $|\bstate{\psi}U_1^\dagger U_2\state{\psi}| \leq 1$, this minimization is equivalent to:
\begin{align}
    P_{\rm s}(\hat{\theta}_C \neq \theta_C)=  \frac{1-\sqrt{1-\min_{\state{\psi}}|\bstate{\psi}U_1^\dagger U_2\state{\psi}|^2}}{2}. 
    \label{eq: error prob U}
\end{align}
Here, $\min_{\state{\psi}}|\bstate{\psi}U_1^\dagger U_2\state{\psi}|$ is nothing but the operator norm $\|U_1^\dagger U_2\|$. Therefore, we look for the bound of $\|U_1^\dagger U_2\|$.

We use a known bound for the operator norm. Consider a $K \times L$ matrix
\begin{align}
    A=
    \begin{pmatrix}
        {\mathbf a}_1 & {\mathbf a}_2 & \cdots & {\mathbf a}_L
    \end{pmatrix},
\end{align}
where ${\mathbf a}_m$ is $K$ dimensional vector. Then the operator norm is bounded as follows. 
For all $l$, 
\begin{align}
    \| A \| \geq \|{\mathbf a}_l\| \label{eq: norm bound}
\end{align}
We apply \eqref{eq: norm bound} to $\|U_1^\dagger U_2\|$ for $d=2$. Then we obtain 
\begin{align}
    P_{\rm s}
    &\geq \frac{1-2\alpha}{2}. 
\end{align}

In the multi-shot protocol, the measurement is performed $m$ times. An estimate $\hat{\theta}_C$ is determined through LRT (Eq.~\ref{eq:lrt_hbcd}) on these measurement outcomes. The probability of error $P_e$ is given by
\begin{align}
    P_e = \frac{1}{2}\left( P_{\hat{\theta}_C|\theta_C}(0|\alpha) + P_{\hat{\theta}_C|\theta_C}(\alpha|0) \right). 
\end{align}
This can be bounded as \cite{cover2005elements}
\begin{equation}
    2^{-m C(p_0, p_\alpha)} \geq P_e \geq \frac{1}{4} \exp \left( -m D(p_0 || p_\alpha) \right),
\end{equation}
where $m$ is the number of measurements, $p_0$ (or $p_\alpha$) is the probability distribution over measurements outcomes $y$ when the truth is $\theta_C = 0$ (or $\theta_C = \alpha$), $C(\cdot)$ is the Chernoff information bound and $D(\cdot)$ is the KL divergence.

To achieve an error probability of at most $\epsilon$, we can then show that
\begin{equation}
    m = O\left( \frac{ \log(1/4\epsilon) } {4 \alpha^2} \right)
\end{equation}
This proves Theorem \ref{thrm: SQL}.

\section{Proof of Theorem \ref{thrm:lower_bound}}\label{app_sec:thrm4}
Our numerical result shows the Heisenberg scaling by the sequential protocol. Here we derive an information-theoretic bound for HBCD and show that the Heisenberg scaling is in fact optimal for $N$ w.r.t. $\alpha$. Let $U_i$ $(i=1,2)$ be a unitary operator in the estimation protocol for $\theta_C=0$ and $\alpha$, respectively. For $U_1$ and $U_2$ to be perfectly discriminated, the following necessary conditions must be satisfied.
\begin{align}
    \mathcal{D}(U_1,U_2)=0,
    \label{eq: distance}
\end{align}
where the distance is defined by 
\begin{align}
    \mathcal{D}(U_1,U_2)=\min_\eta |\bstate{\eta} U_1^\dagger U_2 \state{\eta}|
\end{align} 
and the minimization is done over all pure states \cite{acin}. 
We evaluate \eqref{eq: distance} using the spectral norm of a matrix with the help of its subadditivity:
\begin{align}
	\| A_1 A_2 -B_1B_2 \| \leq \|A_1 -B_1\| + \|A_2 -B_2\|.     
\end{align}
Then, the upper bound of the distance between $U_1$ and $U_2$ is obtained:
\begin{align}
        \|U_1-U_2\| \leq 
    \sum_{i=1}^N \|C(\theta_C=0)-C(\theta_C=\alpha)\|
    = N\sqrt{2(1-\cos \alpha)} \label{eq: norm U1 U2}
.
\end{align}
It quantifies the distance between two unitary operators. Then we translate the distance of operators to the distinguishability of them. 
Substituting the upper bound of the distance into Eq.~\eqref{eq: distance}, we obtain
\begin{align}
    \mathcal{D}(U_1,U_2) \geq 1 - N \sqrt{2(1-\cos \alpha)}.     
\end{align}
Therefore, the necessary condition for \eqref{eq: distance} is 
\begin{align}
    N  \geq {1\over \sqrt{2(1-\cos \alpha)}}, 
\end{align}
i.e., this gives a lower bound of the query complexity for perfect discrimination shown in Fig.~\ref{fig:numerical_bounds_hbcd}. Here we can explicitly see that the bound for $N$ is the Heisenberg limit for small $\alpha$, 
\begin{align}
    N \sim \frac{1}{\alpha}
\end{align} 

\section{Numerical Experiments}\label{app_sec:numerical_experiments}
\begin{figure}[ht!]
    \centering
    \includegraphics[scale=1.5]{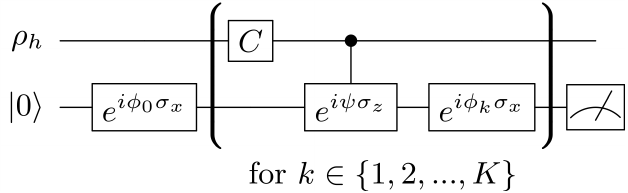}
    \caption{Quantum circuit used in numerical experiments for HBCD using the sequential or multi-shot protocols}
    \label{fig:schematic_hbcd_circuit}
\end{figure}

In our numerical experiments for solving the problem of HBCD using the sequential protocol or multi-shot protocol, we consider the circuit of Figure~\ref{fig:schematic_hbcd_circuit}, shown with a phase sequence of length $K$. Let us describe how this compares to the corresponding circuits described in Figure~\ref{fig:Strategies}. Here, $\rho_h = I/2$ and $\rho_m$ are prepared through the action of the single qubit gate of $R_x(\phi_0)$ on the zero state $\ket{0}$. In a simplification from Figure~\ref{fig:Strategies}(a), we consider the phase of the controlled gate to be the same across multiple applications i.e., $\psi_1=\ldots=\psi_K=\psi$. We then denote the overall phase sequence as $\Phi=\{\phi_0,\phi_1,\ldots,\phi_K,\psi\}$, which now includes $\phi_0$. Note that in the main text, we typically denote the length of the phase sequence $K$ as $N$ for the sequential protocol as we measure only once, and as $d$ for the multi-shot protocol with constant depth queries.

Having described the circuit to be used in our protocols, we are now in a position to describe how a phase sequence $\Phi$ is optimized for the problem of HBCD and further details of our experimental procedures for the sequential and multi-shot protocols.

\paragraph{Optimization of phase sequences}
Given length $K$, we obtain a numerically optimized phase sequence $\hat{\Phi}$ by solving the optimization of Eq.~\ref{eq:optimization_prob_hidden_BCD} using the quasi-Newton method of L-BFGS with a particular choice of initial conditions. Denoting the initial condition as $\Phi^0$, we set its first $K+1$ components as
\begin{equation}
    \Phi^0_{1:K+1} = \left \{\phi^0_0,\phi^0_1,\ldots,\phi^0_{K-1},\phi^0_K \right\} = \left \{\frac{\pi}{4},0,\ldots,0,\frac{\pi}{4} \right\}.
    \label{eq:initial_conditions_phase_seq}
\end{equation}
This choice of initial conditions for the phases was inspired by work on optimization of phases in quantum signal processing \cite{dmwl_21,wang2022energy}, using gradient-based methods. The last component $\Phi^0_{K+2}=\psi^0$ is set randomly by sampling from a normal distribution with zero mean and unit variance. We prepare $n_{reps}$ such initial conditions and then run the L-BFGS algorithm to solve Eq.~\ref{eq:optimization_prob_hidden_BCD}. This results in $n_{reps}$ different phase sequence solutions from which we select the one with the lowest loss. In our numerical experiments on HBCD with the sequential and multi-shot protocols, we found that choosing $n_{reps}=10$, this choice of initial conditions and optimization method yielded solutions at the global minima of the loss function in Eq.~\ref{eq:optimization_prob_hidden_BCD}.

\begin{figure}[ht!]
    \centering
    \includegraphics[scale=0.5]{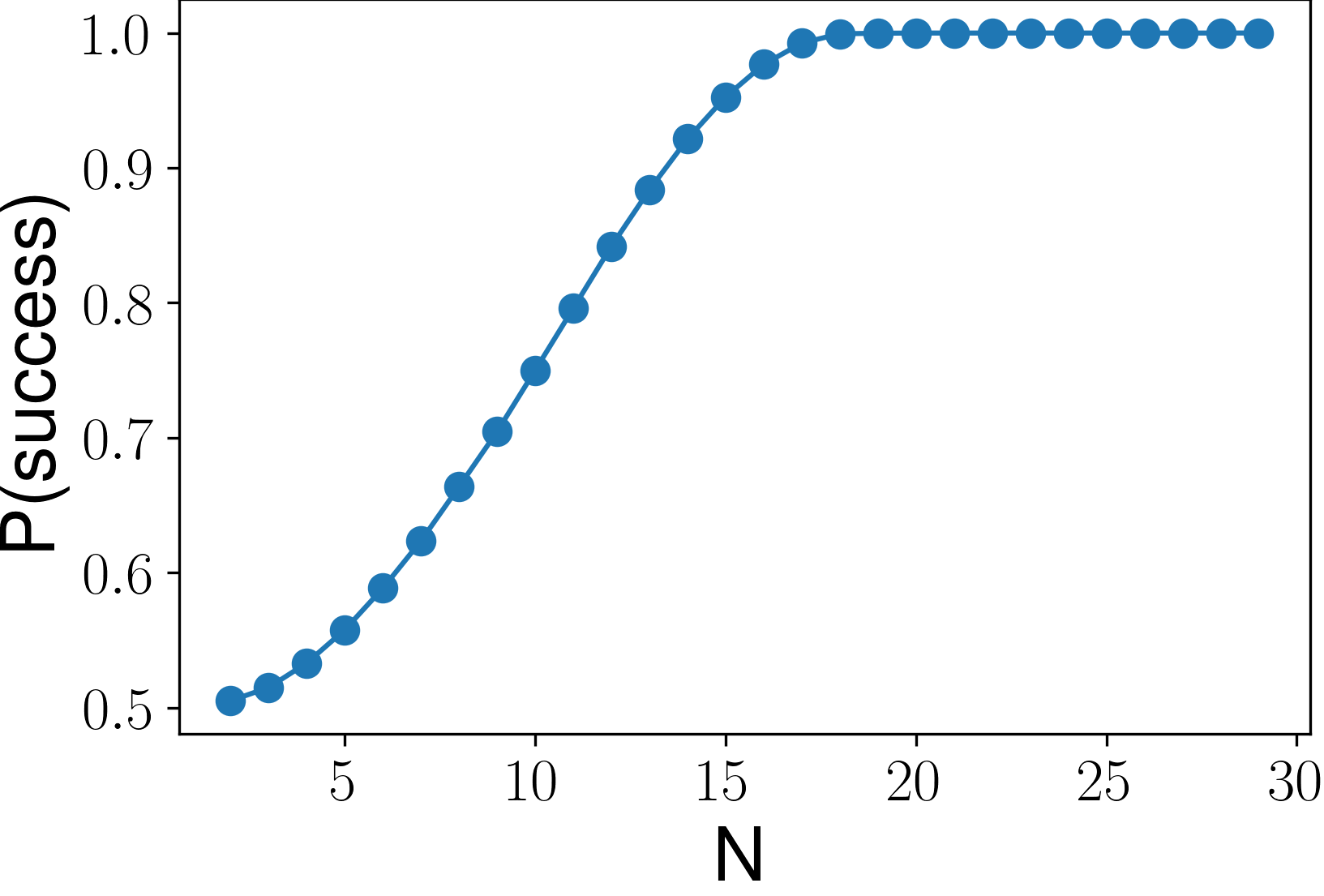}
    \caption{Probability of success of the sequential protocol in HBCD of $\theta_C=\alpha=0.1$ from $\theta_C=0$ with increasing length $N$ of phase sequence $\Phi$.}
    \label{fig:prob_success_seq_protocol}
\end{figure}

\paragraph{Sequential protocol}
In our numerical experiments with the sequential protocol, we assume that we only measure once. The goal is to then determine the length $N$ of the phase sequence $\Phi$ at which we are able to discriminate $\theta_C=\alpha$ from $\theta_C=0$. As described in the main text, we do this by starting at $N=1$ and increasing the value of $N$ by one until we determine a numerically optimized phase sequence $\Phi$ (through the optimization procedure described above) that yields a probability of error less than equal to a given error parameter $\epsilon \in [0,1/2)$ or probability of success greater than equal to $1-\epsilon$. In Figure~\ref{fig:prob_success_seq_protocol}, we show how the probability of success varies with $N$ for $\alpha=0.1$. As illustrated in the figure, the probability of success increases with $N$, reaching a value of $>0.95$ for $N=16$.
\paragraph{Multi-shot protocol} In our numerical experiments with the multi-shot protocol with constant depth $d$ queries, we fix the length of phase sequence to $d$ and measure $m$ times. The goal is to then determine the minimal number of shots $m^\star$ required to solve HBCD of $\theta_C=\alpha$ from $\theta_C=0$ with an error probability below $\epsilon$ with a numerically optimized phase sequence $\Phi$ of length $d$. This needs to be done empirically as an analytical expression of the error probability over $m$ shots is not available to us. We now describe our experimental procedure for determining the value of $m^\star$ for a given value of $\alpha$ and $d$. Such experimental procedures are common in the statistical learning community \cite{lokhov2018optimal}.

For a given value of $\alpha$, we first determine a numerically optimized phase sequence $\Phi$ of length $d$ to solve HBCD using the optimization procedure described earlier. We then generate $L \geq 1$ independent sets (indexed by $t$) of $m$ measurement outcomes by sampling $\theta_C^{t}$ uniformly from $\{0,\alpha\}$ for each set and using $\Phi$. The likelihood ratio test (LRT) (Eq.~\ref{eq:lrt_hbcd}) is then used to determine the estimate $\hat{\theta}_C$ on each of the $L$ sets of measurement outcomes yielding $L$ estimates: $\{\hat{\theta}_C^t\}_{t \in [L]}$. If all the estimates $\hat{\theta}_C^t$ correctly match the corresponding truth $\theta_C^t$ for all $t \in \{1,2,\ldots,L\}$, we say that the multi-shot protocol succeeded in HBCD within a given error probability $\epsilon$.

We now describe how to obtain the value of $L$ given the error probability $\epsilon$, required in our numerical experiments to guarantee that the multi-shot protocol succeeds with a probability above $1-\epsilon$ with confidence at least $95\%$. Let the probability of success on any of the $L$ sets i.e., $P(\hat{\theta}_C^t = \theta_C^t)$ be equal to $p$. Note that the outcome of $\hat{\theta}_C^t$ being equal to $\theta_C^t$ through the course of our numerical experiment is then equivalent to generating flips of an unfair coin with the probability of success equal to $p$. Assuming a uniform initial prior on $p$, let us denote $P_{post}(p | L)$ as the posterior probability over $p$ after a series of $L$ successful estimations, which is given by the Beta distribution for this Bernoulli process. We then have the probability of confidence $p_{conf}$ in the value of $p$ given successive $L_{succ}$ successful estimations as
\begin{equation}
    p_{conf} = \int_{1 - \epsilon}^{1} P_{post}(p | L_{succ} = L) dp.
\end{equation}
We require that $p_{conf} \geq 0.95$, which is obtained first for $L = 59$ for $\epsilon=0.05$, $L=119$ for $\epsilon=0.025$, and $L=598$ for $\epsilon=0.005$. These values of $L$ were used in all our numerical experiments with the multi-shot protocol in this work.

\section{Operating Characteristics}\label{app_sec:operating_characteristics_hbcd}
We have commented so far on the performance of different discrimination protocols (Def.~\ref{def:HBCD_protocol}) in terms of the number of queries $N$ required to solve an HBCD problem $\textsf{HBCD}(\alpha,\epsilon,\rho_h)$ (Def.~\ref{def:HBCD_problem}). In this section of the supplementary material, we will comment on the performance of discrimination protocols in terms of detection probability under constraints on the total number of queries allowed.

In classical binary hypothesis testing \cite{cover2005elements}, the performance of any estimator (or decision rule) can be specified fully in terms of the detection probability $P_D$ and the false-alarm probability $P_F$, defined as follows
\begin{align}
    P_D &= \mathbb{P}\left( \hat{\theta}_C(\mathbf{y}) = \alpha \,\mid \, \theta_C = \alpha \right), \\
    P_F &= \mathbb{P}\left( \hat{\theta}_C(\mathbf{y}) = \alpha \,\mid \, \theta_C = 0 \right).
    \label{eq:prob_detection_failure}
\end{align}
For an estimator, it is desired to have a high value of $P_D$ with a lower value of $P_F$. There may, of course, be additional criteria. For the sequential protocol, we want to achieve $P_D$ higher than a certain value (say $a_D$) while keeping $P_F$ below a certain threshold (say $a_F$) for the minimal length of the phase sequence $\Phi$. In the multishot protocol with constant query depth $d$, we have the same goal but for the minimal number of shots $m$.

In Appendix~\ref{app_sec:estimators}, we noted that the estimator $\hat{\theta}_C(\cdot)$ of choice for both the sequential and multishot protocols is the likelihood ratio test (LRT, Eq.~\ref{eq:lrt_hbcd}). We can rewrite this in the following form for a discrimination protocol $\Sigma$ (Def.~\ref{def:HBCD_protocol})
\begin{equation}
    \hat{\theta}_C = 
    \begin{cases}
    \alpha &, \, \frac{p_{\mathbf{y}|\theta_C}\left( \mathbf{y} | \alpha; \Sigma \right) }{ p_{\mathbf{y}|\theta_C}\left( \mathbf{y} | 0 ; \Sigma \right) } \geq \eta, \\
    0 &, \, \frac{p_{\mathbf{y}|\theta_C}\left( \mathbf{y} | \alpha ; \Sigma \right) }{ p_{\mathbf{y}|\theta_C}\left( \mathbf{y} | 0 ; \Sigma \right) } < \eta,
    \end{cases}
    \label{eq:lrt_threshold}
\end{equation}
where $\eta \in [0,\infty)$ is some threshold determined by the choice of prior probabilities and cost criterion. Formerly in Eq.~\ref{eq:lrt_hbcd}, this threshold had the value of $1$ as we considered the prior probabilities of $\theta_C$ to be uniform and the cost of making any decision to have equal risk. We note that specifying the value of $\eta$ specifies the decision rule in Eq.~\ref{eq:lrt_threshold}. In fact, we can describe the decision regions in terms of the measurement outcomes $\mathbf{y} \in \{0,1\}^m$ as follows:
\begin{align}
    \mathcal{D}(\eta) &= \{ \mathbf{y} : p_{\mathbf{y}|\theta_C}\left( \mathbf{y} | \alpha; \Sigma \right) / p_{\mathbf{y}|\theta_C}\left( \mathbf{y} | 0 ; \Sigma \right) \geq \eta \}, \\
    \bar{\mathcal{D}}(\eta) &= \{ \mathbf{y} : p_{\mathbf{y}|\theta_C}\left( \mathbf{y} | \alpha; \Sigma \right) / p_{\mathbf{y}|\theta_C}\left( \mathbf{y} | 0 ; \Sigma \right) < \eta \},
    \label{eq:lrt_decision_regions}
\end{align}
where $\mathcal{D}$ denotes the set of measurement outcomes $\mathbf{y}$ for which the LRT makes the decision of $\hat{\theta}_C = \alpha$ and $\bar{\mathcal{D}}$ denotes the complement of $\mathcal{D}$ or the set of measurement outcomes for which the LRT makes the decision of $\hat{\theta}_C=0$. 

Moreover, it is known that the LRT maximizes the detection probability for a given upper bound on the false-alarm probability i.e., LRT is optimal under the Neyman-Pearson criterion \cite{helstrom1994elements}. Each value of $\eta$ can thus be associated with a particular $(P_D,P_F)$ operating point. Varying the value of $\eta$ varies the decision regions (Eq.~\ref{eq:lrt_decision_regions}) and thus allows us to analyze the trade-off between detection probability and false-alarm probability. The resulting curve of $(P_D,P_F)$ points from varying $\eta \in [0,\infty)$ is called the operating characteristic.

We now adapt the operating characteristics to our quantum setting of HBCD in a similar fashion to that of quantum detector operating characteristics (QDOC) in \cite{medlock2019operating} and will also call them by the same name. For a given discrimination protocol $\Sigma(N,d,Z,S)$ (see Def.~\ref{def:HBCD_protocol} for details on inputs) which includes specifying the type of protocol $S$ and phase sequence $\Phi$, we generate QDOC by varying the decision regions (Eq.~\ref{eq:lrt_decision_regions}) i.e., the value of $\eta$ in LRT. Note that the phase sequence $\Phi$ and total number of queries $N$ used are then fixed a priori. Let us now analyze QDOCs for the sequential and multi-shot protocols in solving HBCD for $\alpha=0.1$ in different scenarios.

\begin{figure}[h!]
    \centering
    \includegraphics[width=0.4\textwidth]{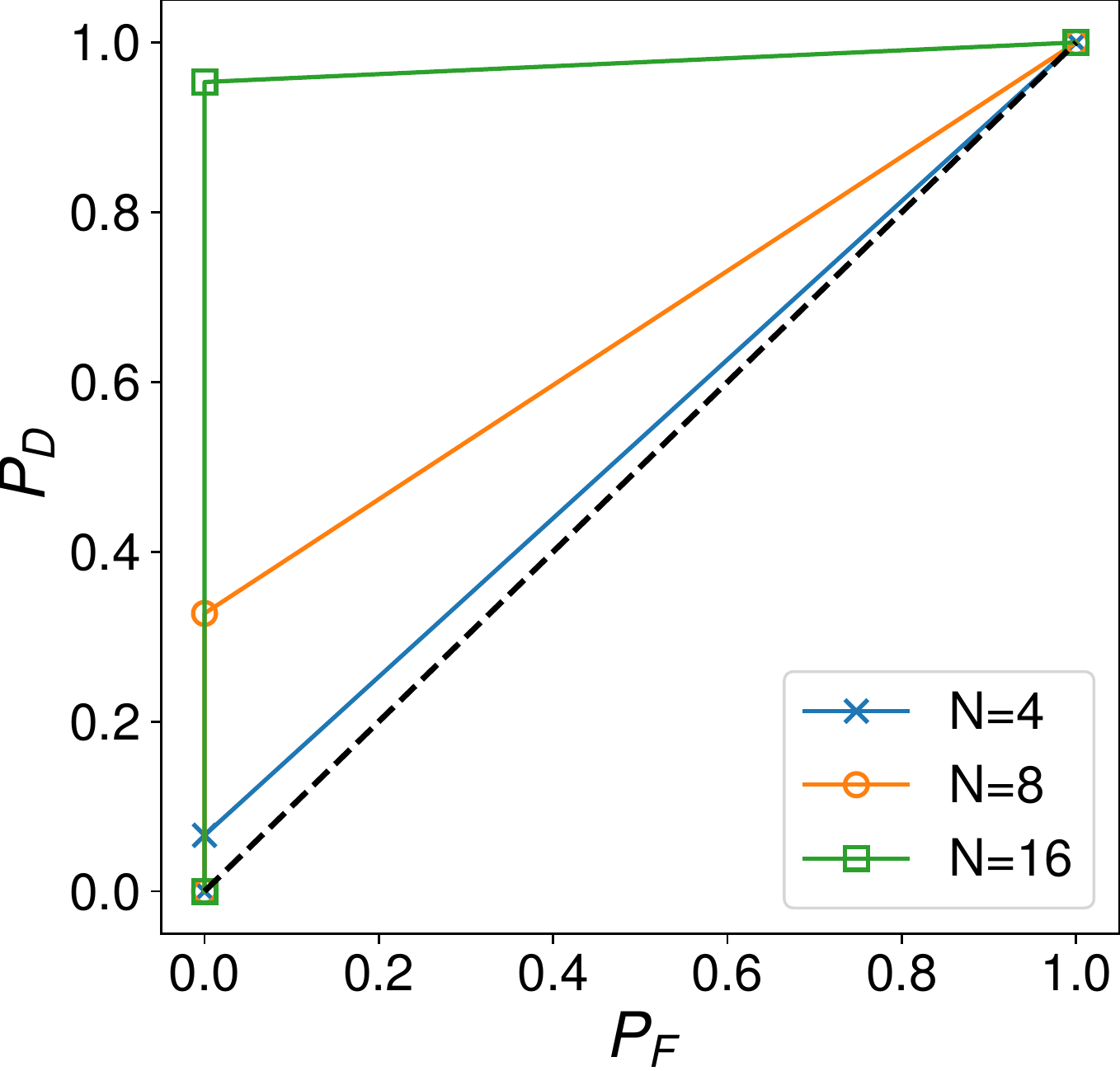}
    \caption{Quantum detector operating characteristics (QDOC) of the sequential protocol in HBCD of $\theta_C=\alpha=0.1$ from $\theta_C=0$ with increasing length $N$ of phase sequence $\Phi$.}
    \label{fig:roc_seq_protocol}
\end{figure}

\textit{Sequential protocol}: In Figure~\ref{fig:roc_seq_protocol}, we plot QDOCs for the sequential protocol in solving HBCD for $\alpha=0.1$ for increasing values of $N$. We observe the QDOCs are monotonically increasing with $P_F$ as expected. There is only one intermediate point of $(P_D,P_F)$ in between $(0,0)$ and $(1,1)$ as we only measure once in the sequential protocol. This operating point corresponds to the value of $\eta=1$. Values of operating points along the piece-wise linear segments can be obtained through randomization \cite{cover2005elements}. In Figure~\ref{fig:roc_seq_protocol}, as the number of concatenated queries $N$ is increased, the detection probability increases reaching $P_D > 0.95$ for $N=16$. This is expected but somewhat surprisingly, we obtain this higher detection probability at a negligible increase in false-alarm probability.

\begin{figure}[h!]
    \centering
    \includegraphics[width=0.85\textwidth]{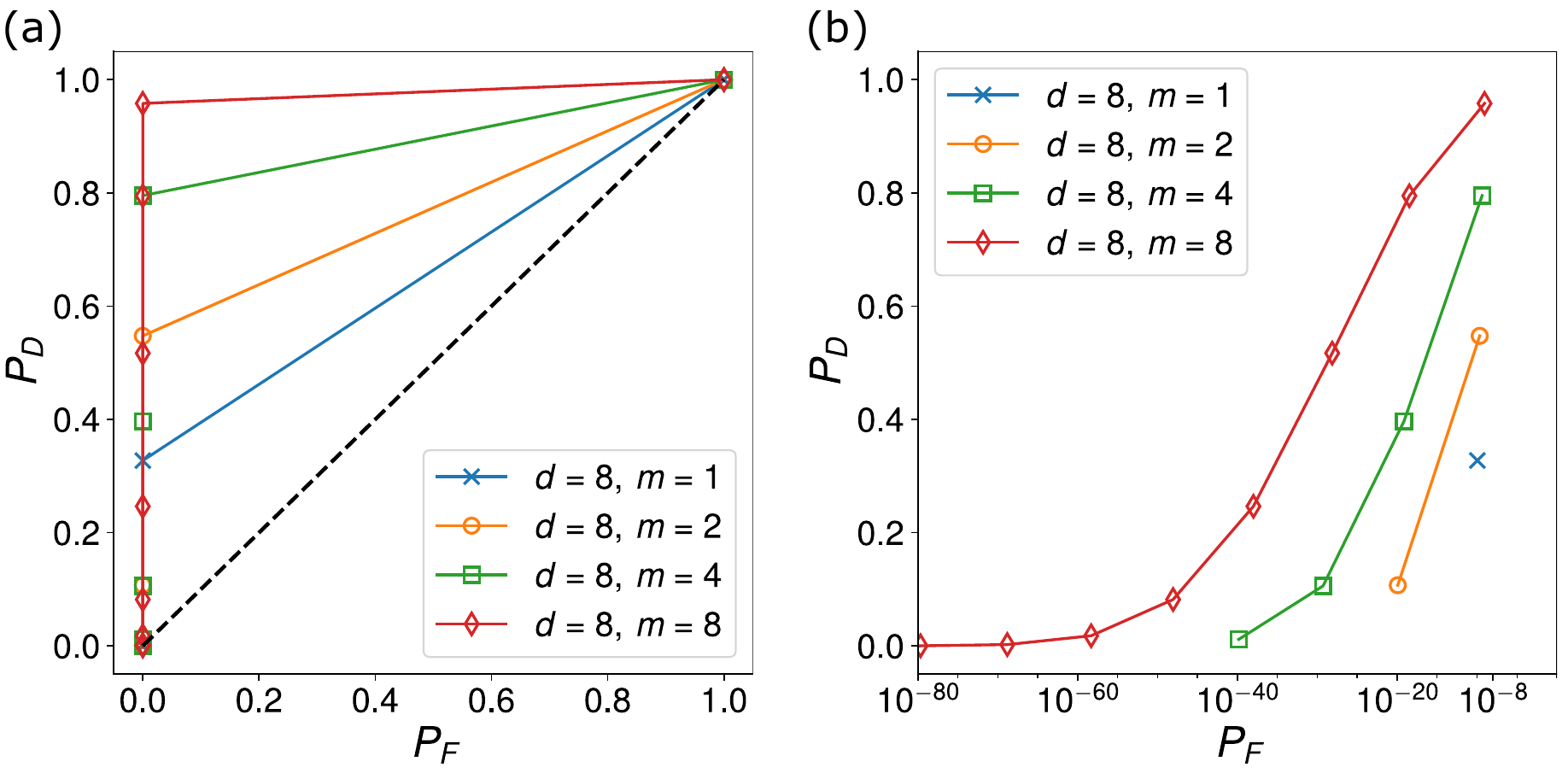}
    \caption{Quantum detector operating characteristics (QDOC) of the multi-shot protocol with a fixed query depth of $d=8$ in HBCD of $\theta_C=\alpha=0.1$ from $\theta_C=0$ with increasing number of shots $m$. (a) Trend of $P_D$ with $P_F$ considering linear scales on both x-axis and y-axis. (b) Trend of $P_D$ with $P_F$ for intermediate operating points obtained by each protocol considering a log-scale on the x-axis to illustrate that $P_F$ remains orders of magnitude below $1$ for all values of $P_D$.}
    \label{fig:roc_multishot_protocol}
\end{figure}

\textit{Multi-shot protocol} In Figure~\ref{fig:roc_multishot_protocol}, we plot QDOCs for the mutli-shot protocol with queries of constant depth $d=8$ and increasing number of shots $m$. For higher values of $m$, there are more number of intermediate operating points corresponding to values of $\eta = \left(\frac{p_{y|\theta_C}(1|\alpha)}{p_{y|\theta_C}(1|0)}\right)^{Y_1}\left(\frac{p_{y|\theta_C}(0|\alpha)}{p_{y|\theta_C}(0|0)}\right)^{m-Y_1}$ with $Y_1 \in \{0,1,\ldots,m\}$ denoting the number of measurement outcomes being one. Increasing the total number of queries $N$ by increasing $m$ allows higher detection probabilities to be achieved and yet again at negligible increase in false-alarm probability.

\begin{figure}[h!]
    \centering
    \includegraphics[width=0.85\textwidth]{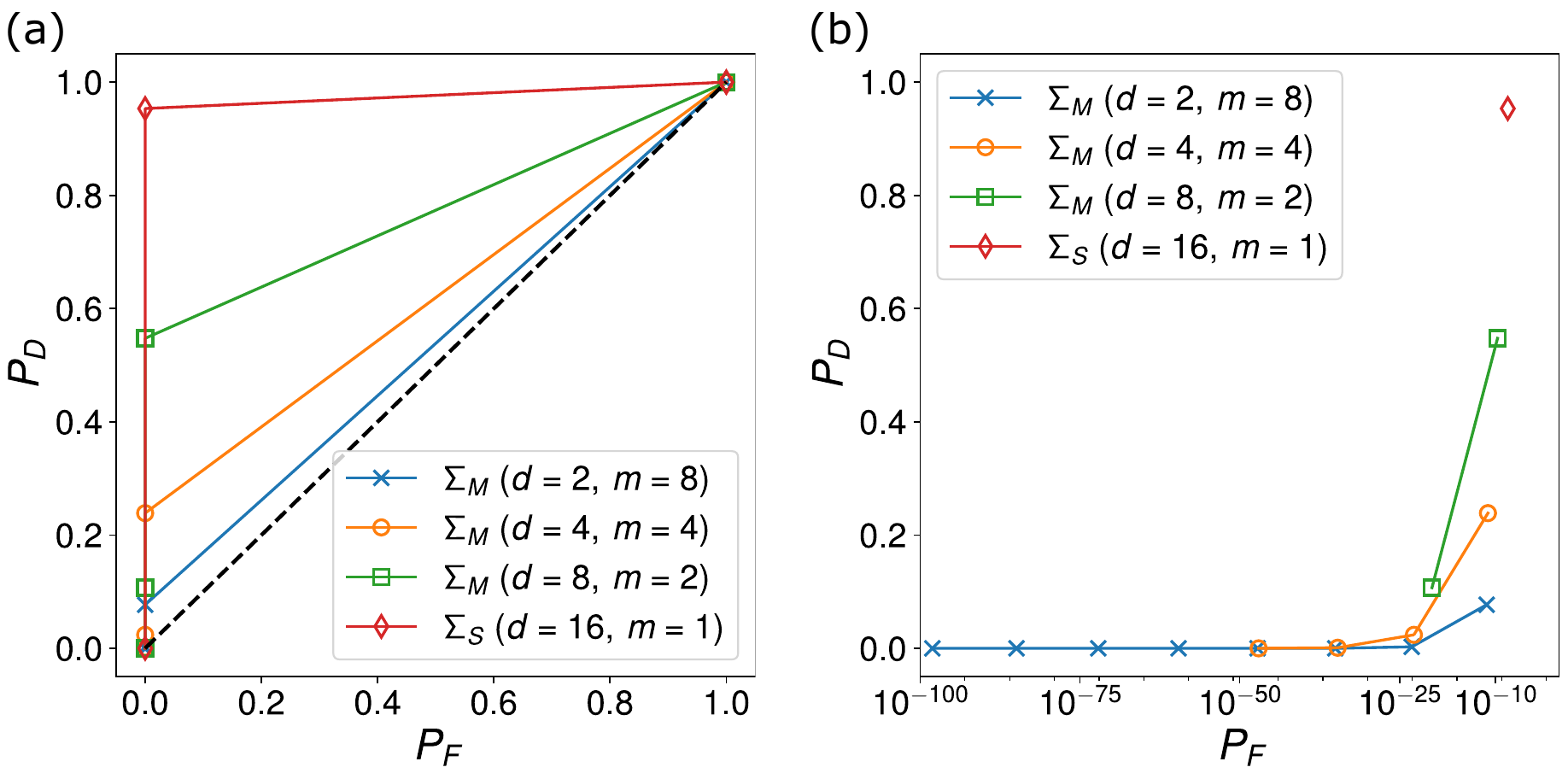}
    \caption{Quantum detector operating characteristics (QDOC) of a sequential protocol ($\Sigma_S$) and multi-shot protocol ($\Sigma_M$) with a fixed budget of $N=16$ in HBCD of $\theta_C=\alpha=0.1$ from $\theta_C=0$. For the multi-shot protocol, we show OCs with increasing query depth $d$ and decreasing number of shots $m$ such that $N=d \cdot m = 16$. (a) Trend of $P_D$ with $P_F$ considering linear scales on both x-axis and y-axis. (b) Trend of $P_D$ with $P_F$ for intermediate operating points obtained by each protocol considering a log-scale on the x-axis to illustrate that $P_F$ remains orders of magnitude below $1$ for all values of $P_D$.}
    \label{fig:roc_fixed_budget_protocol}
\end{figure}

\textit{Fixed resource budget} We now consider the scenario where the total number of queries allowed to be used by any protocol is fixed to $N=16$. In Figure~\ref{fig:roc_fixed_budget_protocol}, we compare the QDOC of the sequential protocol against various multi-shot protocols. We find that even under constraints of experimental resources, it is advantageous to use a sequential protocol to obtain a higher detection probability than any corresponding multi-shot protocol with same resource constraints.
%

\section{Example of HBCD}\label{app_sec:HBCD_example}
In this appendix, we provide a more detailed discussion of the experimental proposal and compare the HBCD protocol to discrimination protocols which use Gaussian states of the photonic degree of freedom.

\subsubsection{On measurement fidelity and single-photon preparation efficiencies:}
While single photons propagating in free space can be considered hidden system primarily due to the difficulty of implementing reconfigurable single qubit gates, they also have disadvantages regarding initialization and measurement. These disadvantages further support their identification as a hidden system and  necessitates the introduction of the measurement system $\mathcal{M}$. First, the efficiency of projective measurements that determine the polarization of a single optical photon is low. Quantum efficiencies of $\approx 95\%$ can rarely be achieved using the most advanced Superconducting Nanowire Based Single-Photon Detectors (SNSPDs) \cite{stasi2022high}.   Secondly, the polarization state of the photon is not easily initialized without significantly reducing the source's quantum efficiency \cite{wang2019towards}. The inefficiency of both the single-photon detector and source increases the cost of the discrimination protocol as the protocol needs to be run multiple times until both procedures succeed \cite{Mosley2008,davis2022improved}. 

\subsubsection{Details of the experimental proposal
} 
The sequential HBCD protocol is implemented in the proposed experiment by allowing a single photon propagate inside a ring cavity and evolve under the influence of linear optical elements as well as the cavity-QED system, which realizes the measurement system $\mathcal{M}$. Hence the number of roundtrips inside the ring cavity corresponds to the depth of the HBCD protocol. We assume that the single photon can be routed in and out of the ring cavity using an electro-optical switch, depicted with a dashed line in Fig. $\ref{fig:Exp}$ (see Ref. \cite{kwiat1999high} for a possible realization). 

We now discuss how each component of the Query $Q$ (see Def.~\ref{def:HBCD_query}) is implemented in the proposed experimental setup during a single roundtrip. The hidden channel is any unitary process that rotates the polarization of the input photon. Next, to implement the required controlled rotation operation, we feed the single photon through a polarizing beam splitter and reflect the $|V\rangle$ arm of the output from the cavity QED system. The interaction between the cavity QED system and the single photon results in a controlled phase gate: $e^{2\psi \ket{1}_{\rm a}\bra{1}\otimes\ket{V}_{p}\bra{V}}$, where the subscripts refer to the atomic (a) and photonic (p) degrees of freedom \cite{duan2004scalable}. 

Here, $e^{i2\psi} \equiv \frac{i\Delta - \kappa }{ i\Delta + \kappa}$, with $\kappa$ and $\Delta$ denoting the cavity decay rate and the detuning of the cavity transition to the optical frequency of the photon, respectively. The controlled phase rotation can be converted to a controlled rotation of the atom along the $z$ axis rotation in the polarization basis by an angle $-\psi$, which can be straightforwardly done by passive linear optical elements when $\psi$ is constant. 
The implementation of the Query is completed with a final $x$ rotation $R_x(\phi_n)$ [see Fig. \ref{fig:Strategies} (a)] on the atom inside the cavity, that needs to be applied before the photon is once again reflected from the cavity-QED system.

\subsubsection{Effect of errors on the HBCD protocol}
Here we show a detailed calculation regarding the error analysis. Let $U_i$ $(i=1,2)$ be a unitary operator of a given sequential protocol for $\theta_C=0$ and $\alpha$, respectively. We compare $U_1$ and $U_2$ to implemented queries $\tilde{U}_1$ and $\tilde{U}_2$, which is defined by
\begin{align}
    \tilde{U}_1 = \tilde{Q}_1^0 \tilde{Q}_2^0 \cdots \tilde{Q}_N^0 
\end{align}
for $\theta_C=0$ and by
\begin{align}
    \tilde{U}_2 = \tilde{Q}_1^\alpha \tilde{Q}_2^\alpha \cdots \tilde{Q}_N^\alpha 
\end{align}
for $\theta_C=\alpha$. Here we define $Q^{x}$ as $Q$ for $\theta_C=x$. 
The error of each query is given by 
\begin{align}
    \|Q^0 - \tilde{Q}^0\| \leq \eta_1
\end{align}
and
\begin{align}
    \|Q^\alpha - \tilde{Q}^\alpha\| \leq \eta_2. 
\end{align}

We analyze the error probability of the single shot in \eqref{eq: error prob U}. 
We replace $U_1$ and $U_2$ with $\tilde{U}_1$ and $\tilde{U}_2$. 
\begin{align}
    P_{\rm s}(\hat{\theta}_C \neq \theta_C)=  \frac{1-\sqrt{1-\min_{\state{\psi}}|\bstate{\psi}\tilde{U}_1^\dagger \tilde{U}_2\state{\psi}|^2}}{2}. 
    \label{eq: error prob U with noise}
\end{align}
We look into the term 
$\min_{\state{\psi}}|\bstate{\psi}\tilde{U}_1^\dagger \tilde{U}_2\state{\psi}|^2$
in the following. 
\begin{align}
    |\bstate{\psi}\tilde{U}_1^\dagger \tilde{U}_2\state{\psi}|
    =&|\bstate{\psi}(\tilde{U}_1-U_1 +U_1)^\dagger (\tilde{U}_2-U_2 +U_2)\state{\psi}| \\
    \leq&|\bstate{\psi}(\tilde{U}_1-U_1)^\dagger (\tilde{U}_2-U_2)\state{\psi}|
    +|\bstate{\psi}U_1^\dagger (\tilde{U}_2-U_2 )\state{\psi}| \nonumber \\
    &+|\bstate{\psi}(\tilde{U}_1-U_1 )^\dagger U_2\state{\psi}|
    +|\bstate{\psi}U_1^\dagger U_2\state{\psi}| \label{eq: error analysis gamma1}\\
    \leq&\|(\tilde{U}_1-U_1)^\dagger (\tilde{U}_2-U_2)\|
    +\|U_1^\dagger (\tilde{U}_2-U_2 )\|
    +\|(\tilde{U}_1-U_1) ^\dagger U_2\|
    +\gamma \label{eq: error analysis gamma2}\\
    \leq&\|(\tilde{U}_1-U_1)^\dagger \| \| (\tilde{U}_2-U_2)\|
    +\|U_1^\dagger \| \| (\tilde{U}_2-U_2 )\|
    +\|(\tilde{U}_1-U_1) ^\dagger \| \|U_2\|
    +\|U_1^\dagger U_2\|\\
    \leq & N^2 \eta_1 \eta_2 + N\eta_2+ N\eta_1 + \gamma,\label{eq: error analysis eta gamma}
\end{align}
where
\begin{align}
    \gamma = \min_{\state{\psi}}|\bstate{\psi}{U}_1^\dagger {U}_2\state{\psi}|. \label{eq: error analysis gamma3},
\end{align}
and the subadditivity of errors entails $\|(\tilde{U}_i-U_i)\| \leq N \eta_i$ for a sequential protocol that consists of $N$ steps in \eqref{eq: error analysis eta gamma}. 
In \eqref{eq: error analysis gamma2}, we evaluate \eqref{eq: error analysis gamma1} with $\state{\psi}$ that minimizes \eqref{eq: error analysis gamma3}, which gives $\gamma$ from the last term of \eqref{eq: error analysis gamma1} and other terms are upper-bounded by the spectrum norm. 

Using \eqref{eq: error analysis eta gamma} and assuming $\eta_1 = O(\eta)$ and $\eta_2 = O(\eta)$ with $N\eta \ll 1$, we evaluate \eqref{eq: error prob U}. 
\begin{align}
    P_{\rm s}(\hat{\theta}_C \neq \theta_C)
    &\leq  \frac{1-\sqrt{1-(N^2 \eta_1 \eta_2 + N \eta_2 + N \eta_1 + \gamma)^2}}{2} \\
    &= \frac{1-\sqrt{1-\gamma^2}}{2}+\frac{2N\gamma (\eta_1 + \eta_2) }{4\sqrt{1-\gamma^2}}+\frac{N^2\{(\eta_1 + \eta_2)^2 + 2 \gamma \eta_1  \eta_2\} }{4\sqrt{1-\gamma^2}} + O(\eta^3)
    \label{eq: error prob gamma eta}
\end{align}
The first term of \eqref{eq: error prob gamma eta} is error probability without implementation errors $\eta_1$ and $\eta_2$. Therefore the leading order of error is the second term, which is $O(N\eta)$. 

However, we can further reduce the error. When the original protocol $U_1$ and $U_2$ achieve perfect discrimination, then $\gamma=0$, and the leading order error is $O(N^2\eta^2)$. 

In the proposed experiment the most dominant error mechanism is due to the gate infidelity of the controlled phase operation discussed in the previous subsection. The major source of errors is the spontaneous emission of a photon from the atom inside optical cavity, with the probability of spontaneous emission averaged over product states $\bar{P}_{\rm se}\approx 1/(4+2C_{\mathrm{QED}})$, where $C_{\mathrm{QED}}$ is the cooperativity of the coupling between the atom and the cavity \cite{duan2004scalable}. Hence, in the following we set $\eta = \sqrt{\bar{P}_{\rm se}}$. 

We finally, make the connection between the operating characteristics and the single-shot error probability $P_s$. We are especially interested in the regime where $P_F$ is negligible. In this regime $P_s \approx P(\theta_C=\alpha)[1-P_D] =  \frac{1}{2}[1-P_D]$ or
\begin{align}
P_D = 1-2P_s
\end{align}
Using this approximation, and the expression for $P_s$, we can calculate $P_D$ as a function of the cooperativity, for a given depth $N$ of the protocol. In order to achieve an advantage over the multi-shot protocol with 8 queries using a single-shot sequential protocol of depth 8, we need to have achieve $P_D\approx 0.2$, which requires $C_{QED} \approx 200$. The protocol can distinguish between different channels when polarization rotation angle $\alpha>0.25$.

\begin{figure}[h!]
    \centering
    \includegraphics[width=0.4\textwidth]{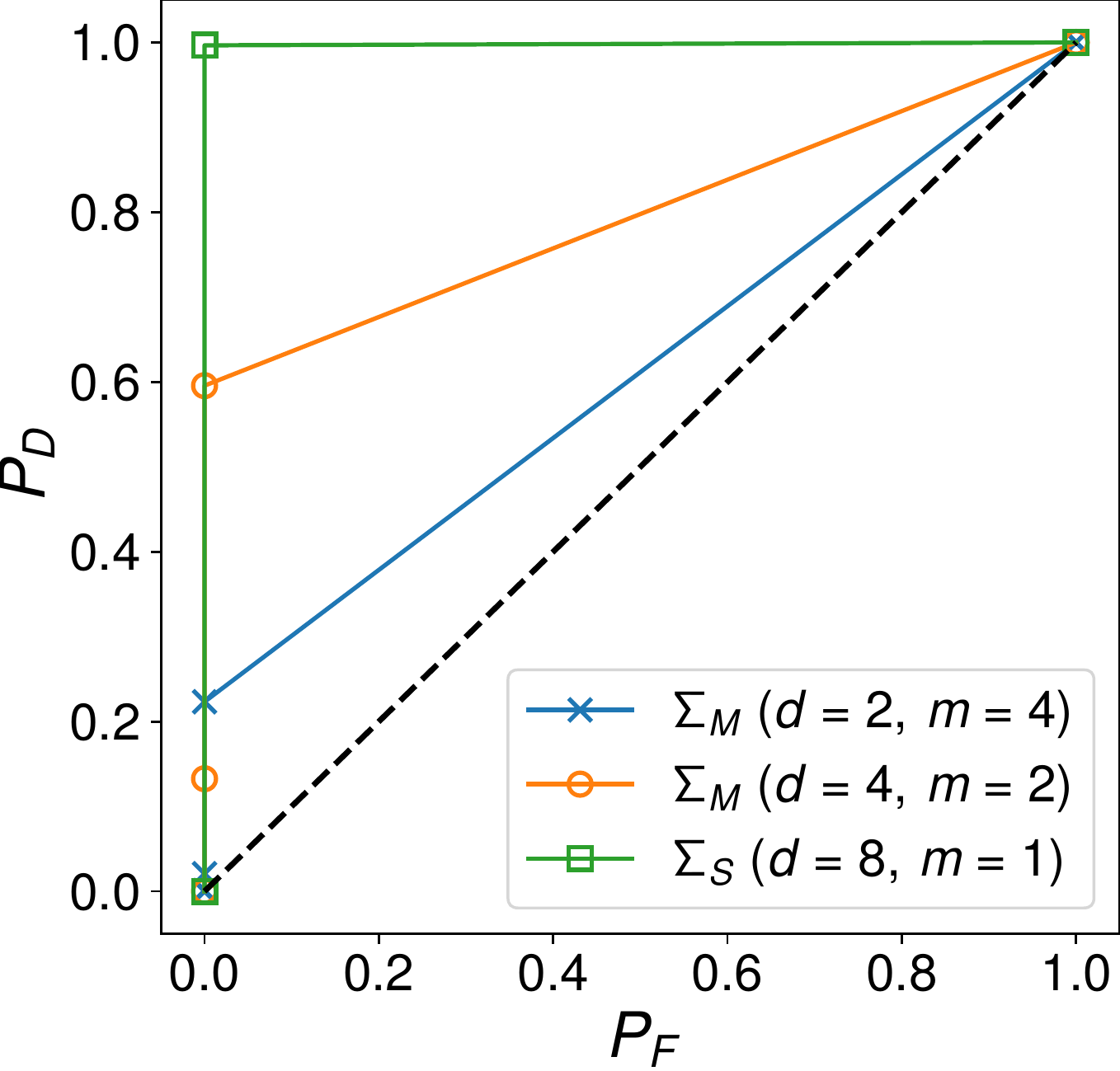}
    \caption{Quantum detector operating characteristics (QDOC) of a sequential protocol ($\Sigma_S$) and multi-shot protocol ($\Sigma_M$) with a fixed budget of $N=8$ in HBCD of $\theta_C=\alpha=0.25$ from $\theta_C=0$. For the multi-shot protocol, we show OCs with increasing query depth $d$ and decreasing number of shots $m$ such that $N=d \cdot m = 8$.}
    \label{fig:roc_fixed_budget_experiment}
\end{figure}

\subsubsection{Comparison to measurements using Gaussian states:} The proposed realization of the HBCD protocol uses only single photon polarization qubits. However, given access to a photonic degree of freedom, the task of distinguishing between the presence or absence of a birefringent slab can also be achieved using Gaussian photonic states which may be easier to prepare using available physical devices. Here, we consider the use of coherent and squeezed states of the photonic degree of freedom and contrast the performance of such discrimination protocols to the one presented in the main text.

First, consider the situation where we detect the presence of the birefringent slab using coherent states, with a fixed linear polarization and a mean number $n_{\rm ph}$ of photons. Whether the birefringent slab is present can be detected by feeding the birefringent slab with such a coherent state and measuring the polarization of the output using a polarizing beam splitter.
However, the variance of the estimate of $\alpha$ in such a protocol scales according to the standard quantum limit with respect to the number of photons used \cite{giovannetti_2011} (~i.e.,~$\propto 1/\sqrt{n_{\rm ph}}$ ~), which is inferior to that obtained with the HBCD protocol.

On the other hand, if we use a combination of  coherent and squeezed-vacuum states \cite{nielsen_chuang} with a, the variance of the estimate scales with the Heisenberg limit \cite{pezze2008mach} with respect to the total number of photons used $n_{\rm ph}$. However, the squeezed vacuum, characterized by the squeezing parameter $r$ is required to satisfy $\sin^2{(r)} = n_{\rm ph}/2$, requiring the preparation of highly squeezed pure states. On the other hand, our protocol does not depend on the purity of the initial photonic state. Moreover, while squeezed vacuum states with $r\approx $ 15 dB has been realized \cite{vahlbruch2016detection}, which only corresponds to about $r= - \ln{(10^{-15/10})}/2\approx 1.7$, and the mean number of photons $\approx 7.5$. Thus the equivalent sequential protocol has depth $<10$.

\end{document}